\documentclass[12pt]{article}

\usepackage[a4paper, total={6.5in, 8.5in}]{geometry}

\usepackage{authblk}

\usepackage{appendix}
\usepackage{algorithmic}
\usepackage[ruled]{algorithm}
\usepackage{pdfpages}

\usepackage{amssymb,amsmath}
\usepackage{verbatim,enumerate}
\usepackage{verbatim,multicol,color}
\usepackage{multirow}
\usepackage{booktabs}
\usepackage{multicol}
\usepackage[english]{babel}
\usepackage{blindtext}
\usepackage{subfig}
\usepackage{float}
\usepackage{bm}
\usepackage{appendix}
\usepackage{xspace}

\usepackage{url}

\usepackage{listings}
\usepackage{color}

\definecolor{dkgreen}{rgb}{0,0.6,0}
\definecolor{gray}{rgb}{0.5,0.5,0.5}
\definecolor{mauve}{rgb}{0.58,0,0.82}

\usepackage{graphicx,graphics}

\def\wh{\widehat}

\def\btheta{{\boldsymbol \theta}}

\providecommand{\e}[1]{\ensuremath{\times 10^{#1}}}

\newtheorem{rem}{Remark}[section]

\newcommand{\bB}{\mathitbf{B}}

\newcommand{\bI}{\mathitbf{I}}

\newcommand{\bZ}{\mathitbf{Z}}
\newcommand{\bA}{\mathitbf{A}}
\newcommand{\bC}{\mathitbf{C}}
\newcommand{\bM}{\mathitbf{M}}

\newcommand{\bL}{\mathitbf{L}}

\newcommand{\bE}{\mathbf{E}}

\newcommand{\bz}{\mathitbf{z}}
\newcommand{\bh}{\mathitbf{h}}
\newcommand{\bs}{\mathbf{s}}

\newcommand{\bv}{\mathbf{v}}

\newcommand{\bmu}{\bm{\mu}}

\newtheorem{defi}{Definition}[section]

\def\supp{\mathop{\rm supp}\nolimits}

\def\cov{\mathop{\rm cov}\nolimits}

\def\diam{\mbox{diam}}
\def\dist{\mbox{dist}}
\def\tr{\mbox{tr}}

\newcommand{\mydet}[1]{\vert #1 \vert}

\DeclareMathAlphabet{\mathitbf}{OML}{cmm}{b}{it}
\def\thetab{\bm{\theta}}
\def\xib{\bm{\xi}}

\def\LL{\mathcal{L}}

\def\bh{\mathbf{h}}
\def\I{\mathitbf{I}}

\def\C{\mathitbf{C}}
\def\L{\mathitbf{L}}

\def\I{\mathitbf{I}}

\def\B{\mathitbf{B}}
\def\A{\mathitbf{A}}

\def\R{\mathitbf{R}}
\def\V{\mathitbf{V}}

\def\U{\mathitbf{U}}
\def\H{\mathcal{H}}
\newcommand{\mcH}{\ensuremath{\mathcal{H}}\xspace}
\newcommand{\landau}[1]{\ensuremath{\mathcal{O}\left(#1\right)}\xspace}

\def\RR{\mathbb{R}}

\def\jmath{j}

\title{HLIBCov: Parallel hierarchical matrix approximation of large covariance matrices and likelihoods with applications in parameter identification}

\author[1]{Alexander Litvinenko  \thanks{Corresponding author}}
\author[2]{Ronald Kriemann}
\author[3]{Marc G. Genton}
\author[4]{Ying Sun}
\author[5]{David E. Keyes}

\affil[1]{RWTH Aachen, Kackertstr. 9C, 52072 Aachen, Germany\\ e-mail: litvinenko@uq.rwth-aachen.de}
\affil[2]{Max-Planck-Institut f\"ur Mathematik in den Naturwissenschaften, 
Inselstr 22, 
04103 Leipzig,
Germany\\e-mail: rok@mis.mpg.de}
\affil[3]{King Abdullah University of Science and Technology, 23955-6900, Thuwal, Saudi Arabia\\ e-mail: marc.genton@kaust.edu.sa}
\affil[4]{King Abdullah University of Science and Technology, 23955-6900, Thuwal, Saudi Arabia\\ e-mail: ying.sun@kaust.edu.sa}
\affil[5]{King Abdullah University of Science and Technology, 23955-6900, Thuwal, Saudi Arabia\\ e-mail: david.keyes@kaust.edu.sa}
\begin{document}
\maketitle
\begin{abstract}
We provide more technical details about the HLIBCov package, which is using parallel hierarchical ($\H$-) matrices to identify unknown parameters of the
covariance function (variance, smoothness, and covariance length). These parameters are estimated by maximizing the joint
Gaussian log-likelihood function. The HLIBCov package
approximates large dense inhomogeneous covariance matrices with a log-linear computational cost and storage requirement.
We explain how to compute the Cholesky factorization, determinant, inverse and quadratic form in the H-matrix format. To demonstrate the numerical performance, we identify three unknown parameters in an example with 2,000,000 locations on a PC-desktop.
\end{abstract}

\maketitle
\tableofcontents

%



\section{Technical details}
\begin{small}
\begin{description}
\item[Program title:] HLIBCov


\item[Nature of problem:]
To approximate large covariance matrices. To perform efficient linear algebra with large covariance matrices on a non-tensor grid.
To estimate the unknown parameters (variance, smoothness parameter, and covariance length) of a covariance function by maximizing the joint Gaussian log-likelihood function with a log-linear computational cost and storage.

\item[Software license:] HLIBCov (GPL 2.0), HLIBpro (proprietary)

\item[CiCP scientific software URL:]

\item[Distribution format:] *.cc files via github

\item[Programming language(s):]  C++

\item[Computer platform:] any

\item[Operating system:] Linux, MacOSX and MS Windows

\item[Compilers:] standard C++ compilers

\item[RAM:] 4 GB and more (depending on the matrix size)

\item[External routines/libraries:]
HLIBCov requires HLIBpro
and GNU Scientific Library (\url{https://www.gnu.org/software/gsl/}).



\item[Running time:] $\mathcal{O}(k^2n \log^2 n)/p$ with $p$ number of CPU cores

\item[Restrictions:]  None (similar limitations as HLIBpro) 

\item[Supplementary material and references:] \url{www.HLIBpro.com} and references therein.

\item[Additional Comments:]
HLIBpro is a software library that implements parallel algorithms for hierarchical matrices. It is freely available in binary form for academic purposes. HLIBpro algorithms are designed for one, two, and three - dimensional problems.
\end{description}
\end{small}
%
%

\section{Introduction}
\label{sec:intro}
HLIBpro is a very fast and efficient parallel $\H$-matrices library. This is an auxiliary technical paper, which contains technical details to our previous paper \cite{Litv17Cov}. In \cite{Litv17Cov} we used the gradient-free optimization method to estimate the unknown parameters of a covariance function using HLIB and HLIBpro. 
%
\paragraph{Parameter estimation and problem settings.} We let $n$ be the number of spatial measurements $\bZ$ located
irregularly across a given geographical region at locations $\bs:=\{\bs_1,\ldots,\bs_n\}\in \Bbb{R}^d$, $d\geq 1$. We also let
$\bZ=\{Z(\bs_1),\ldots,Z(\bs_n)\}^\top$, where $Z(\bs)$ is a stationary Gaussian random field. Then, we assume that $\bZ$ has mean
zero and a stationary parametric covariance function $C(\bh;\btheta)=\cov\{Z(\bs),Z(\bs+\bh)\}$, where $\bh\in\Bbb{R}^d$ is a
spatial distance and vector $\btheta\in\Bbb{R}^q$ denotes $q$ unknown parameters. To infer $\btheta$, we maximize the joint Gaussian log-likelihood function,
\begin{equation}
\label{eq:likeli}
\LL(\thetab)=-\frac{n}{2}\log(2\pi) - \frac{1}{2}\log \mydet{\bC(\thetab)}-\frac{1}{2}\bZ^\top \bC(\thetab)^{-1}\bZ,
\end{equation}
where $\bC(\thetab)_{ij}=C(\bs_i-\bs_j;\btheta)$, $i,j=1,\ldots,n$. 
Let us assume that $\wh \btheta$ maximizes (\ref{eq:likeli}). When the sample size $n$ is large, the evaluation of
(\ref{eq:likeli}) becomes challenging, due to $\cal{O}(n^3)$ computational cost of the Cholesky factorization. Hence, scalable and efficient methods that can process larger $n$ are
needed.

For this, the hierarchical matrices (\mcH-matrix) technique is used, which approximates sub-blocks of the dense matrix
by a low-rank representation of either a given rank \(k\) or a given accuracy \(\epsilon > 0\) (see
Section~\ref{sec:Happrox}).
\begin{defi}
An $\H$-matrix approximation with maximal rank $k$ of the exact log-likelihood $\LL(\thetab)$ is defined by $\widetilde \LL(\thetab;k)$:
\begin{equation}
\label{eq:likeH}
\widetilde{\LL}(\thetab;k)=-\frac{n}{2}\log(2\pi) - \sum_{i=1}^n\log \{\widetilde{\bL}_{ii}(\thetab;k)\}-\frac{1}{2}\bv(\thetab)^\top \bv(\thetab),
\end{equation}
where $\widetilde{\bL}(\thetab;k)$ is an $\H$-matrix approximation of the Cholesky factor $\bL(\btheta)$ with 
maximal rank $k$ in the sub-blocks, $\bC(\btheta)=\bL(\btheta)\bL(\btheta)^\top$, and vector $\bv(\thetab) $ is the solution of the system $\widetilde{\bL}(\thetab;k)\bv(\thetab)=\bZ$.
\end{defi}

To maximize $\widetilde{\LL}(\thetab;k)$ in (\ref{eq:likeH}), we use the Brent-Dekker method
\cite{Brent73,BrentMethod07}. It could be used with or without derivatives. 

An additional difficulty is the ill-posedness of the optimization problem. Even a small perturbation in the covariance matrix
$\bC(\btheta)$ may result in large perturbations in the log-determinant and the log-likelihood. A
possible remedy, which may or may not help, is to take a higher rank $k$.
\paragraph{Features of the $\H$-matrix approximation.} Other advantages of applying the $\H$-matrix technique are the following:
\begin{enumerate}
\item The $\H$-matrix class is large, including low-rank and sparse matrix classes;
\item $\bC(\thetab)^{-1}$, $\bC(\thetab)^{1/2}$, $\mydet {\bC(\thetab)}$, Cholesky decomposition, the Schur complement,
  and many others can be computed in the $\H$-matrix format \cite{HackHMEng};
\item Since the $\H$-matrix technique has been well studied, there are many examples, multiple sequential and parallel
  implementations and a solid theory already available. Therefore, no specific MPI or OpenMP knowledge is needed;
\item The $\H$-matrix cost and accuracy is controlled by $k$;
\item The $\H$-Cholesky factor and the $\H$-inverse have moderate ranks.
\end{enumerate}
\paragraph{Structure of the paper.} In Section \ref{sec:methods}, we introduce the $\H$-matrix approximations of Mat\'ern covariance matrices and Gaussian likelihood functions. 
In Section \ref{sec:conv}, we estimate the memory storage and computing costs. In Section~\ref{sec:install},
we describe the software installation details, procedures of the HLIBCov code,
and the algorithm for parameter estimation.
The estimation of unknown parameters is reported in Section~\ref{sec:MC}. Best practices are listed in Section~\ref{sec:Best}.
We end the paper with a conclusion in Section~\ref{sec:Conclusion}. The auxiliary $\H$-matrix details are provided in the Appendix~\ref{App:Appendix}.
\section{Methodology and algorithms}
\label{sec:methods}
\subsection{Mat\'{e}rn covariance functions}
\label{sec:Matern}
Mat\'{e}rn covariance functions \cite{Matern1986a} are very widely used class of functions \cite{Guttorp2006a,Handcock1993a}. 


For any two spatial locations $\bs$ and $\bs'$ and the distance $\bh:=\Vert \bs-\bs'\Vert $,
the Mat\'{e}rn class of covariance functions is defined as
\begin{equation}
\label{eq:MaternCov}
C(\bh;\btheta)=\frac{\sigma^2}{2^{\nu-1}\Gamma(\nu)}\left(\frac{\bh}{\ell}\right)^\nu \mathcal{K}_\nu\left(\frac{\bh}{\ell}\right),
\end{equation}
where $\btheta=(\ell,\nu,\sigma^2)^\top$; $\ell>0$ is a spatial range parameter; $\nu>0$ is the smoothness, with larger values of $\nu$ corresponding to smoother random fields; and $\sigma^2$ is the variance. Here, ${\cal K}_\nu$ denotes a modified Bessel function of the second kind of order $\nu$, and $\Gamma(\cdot)$ denotes the Gamma function. 
The values $\nu=1/2$ and $\nu=\infty$ correspond to the exponential and Gaussian covariance functions respectively.
%
%
%
%
%
\subsection{Introduction to hierarchical matrices}
\label{sec:Happrox}
Detailed descriptions of hierarchical matrices \cite{HackHMEng, Part1, Part2, GH03, weak, MYPHD} and their applications can be found elsewhere \cite{khoromskij2008data, BoermGarcke2007, harbrecht2015efficient, ambikasaran2013large, ambikasaran2014fast, litvinenko2009sparse, khoromskij2009application}.

The $\H$-matrix technique was originally introduced by Hackbusch (1999) for the approximation of stiffness matrices
and their inverses coming from partial differential and integral equations \cite{Part1, GH03, Winter}. Briefly, the key
idea of the $\H$-matrix technique is to divide the initial matrix into sub-blocks in a specific way, identify those
sub-blocks which can be approximated by low-rank matrices and compute the corresponding low-rank approximations.

\begin{figure}[htbp!]
\centering
\includegraphics[width=0.4\textwidth]{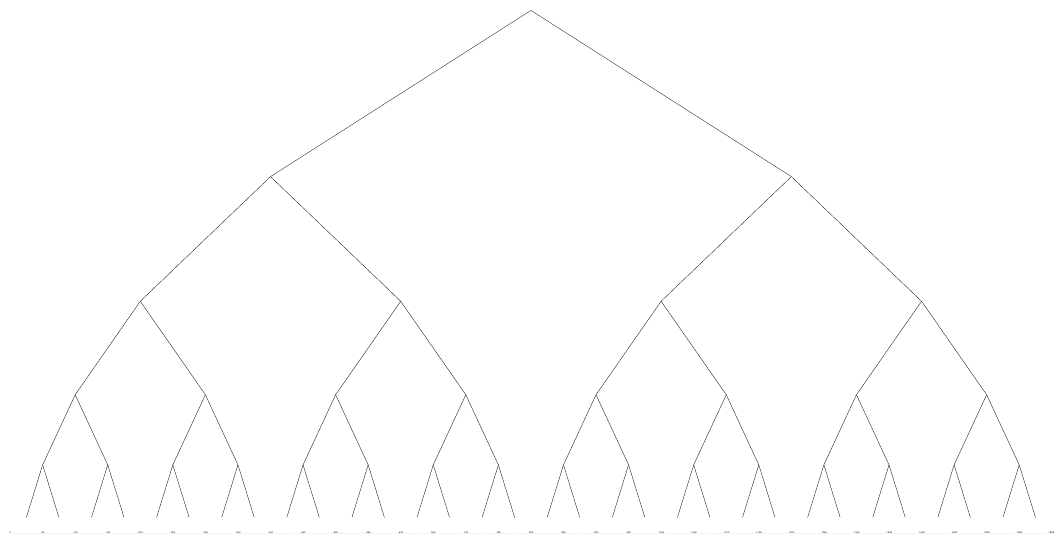}
\includegraphics[width=0.58\textwidth]{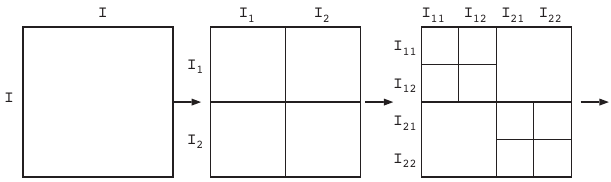}
\caption{Examples of a cluster tree $T_I$ (left) and a block cluster tree $T_{I\times I}$ (right). 
The decomposition of the matrix into sub-blocks is defined by $T_{I\times I}$ and the admissibility condition.}
\label{fig:block_ct}
\end{figure}

The partitioning of the matrix into sub-blocks starts by recursively dividing the rows and columns into disjoint
sub-sets, e.g., splitting the set of all rows into two (equal sized) sub-sets, which are again divided. This
yields a \emph{cluster tree} where each sub-set of rows/columns is called a \emph{cluster}. By multiplying the cluster
trees for the rows and the columns, a hierarchical partitioning of the matrix index set is obtained, the so called
\emph{block cluster tree} or \emph{\mcH-tree}. Within this block cluster tree, low-rank approximable blocks are
identified using an \emph{admissibility condition}. Such \emph{admissible} blocks are not further refined into
sub-blocks, i.e., the corresponding sub-tree is not computed or stored. For all admissible blocks a low-rank
approximation of the initial matrix is computed, either with a given rank \(k\) (fixed-rank strategy) or an accuracy
\(\varepsilon > 0\) (fixed-accuracy strategy). The result of this computation is called an \mcH-matrix. This process is
also shown in Figure~\ref{fig:block_ct}.

\begin{defi}
\label{def:Hmatrix}
Let $I$ be an index set (representing the rows/columns) and $T_I$ be a cluster tree based in \(I\). Furthermore let
$T_{I\times I}$ be a block cluster tree based on \(T_I\) and an admissibility condition
\(\operatorname{adm} : T_{I\times I} \rightarrow \{\mathnormal{true},\mathnormal{false}\}\). Then the set of
\mcH-matrices with maximal rank \(k\) is defined as
\begin{equation*}
\mathcal{H}(T_{I \times I},k):=\{\bC \in \mathbb{R}^{I \times I}\, \vert \, \operatorname{rank}(\bC\vert_{t \times s}) \leq k
\text{ for all } (t,s) \text{ of } T_{I\times I} \text{ with } \operatorname{adm}(t,s) = \text{true} \}.
\end{equation*}
\end{defi} 

\begin{figure}[htbp!]
\centering
\includegraphics[width=0.3\textwidth]{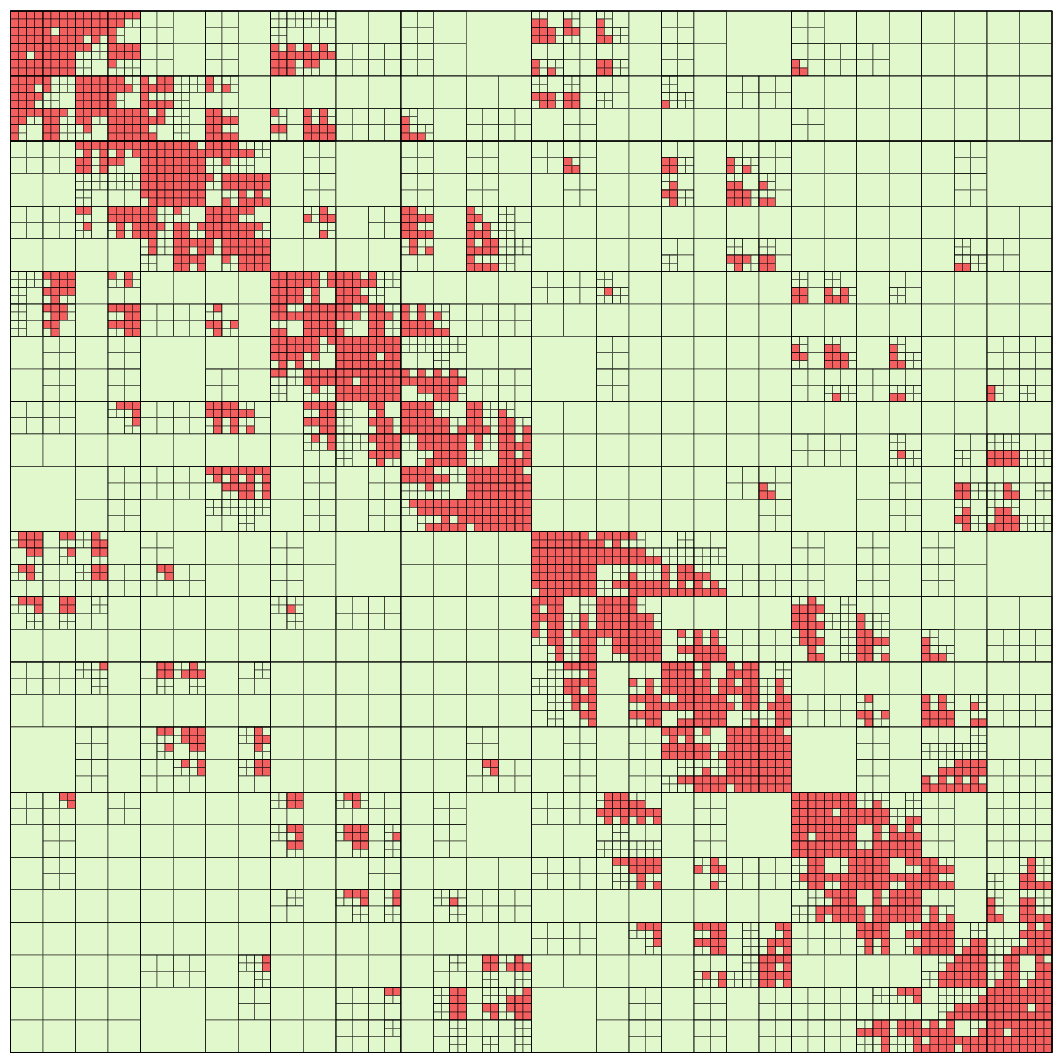}
\includegraphics[width=0.3\textwidth]{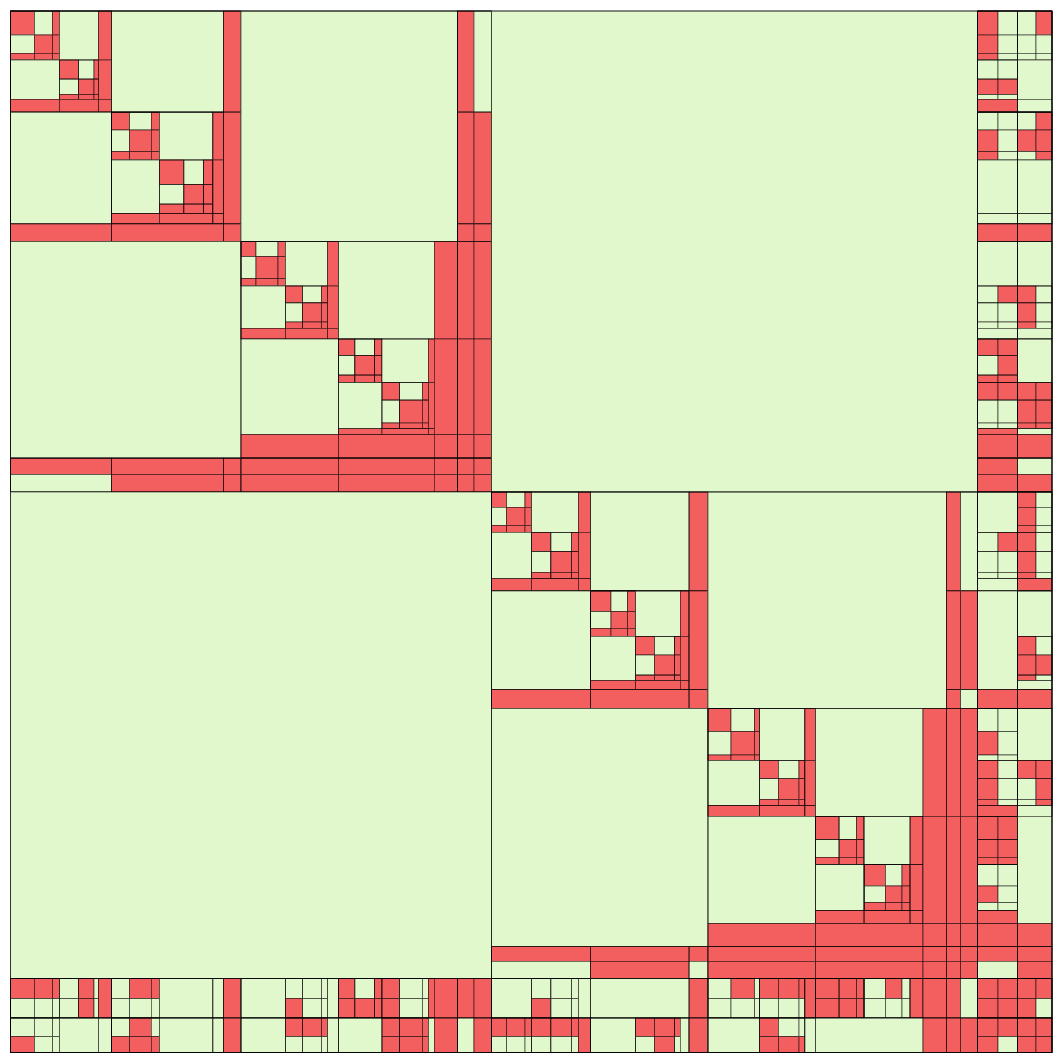}
\includegraphics[width=0.3\textwidth]{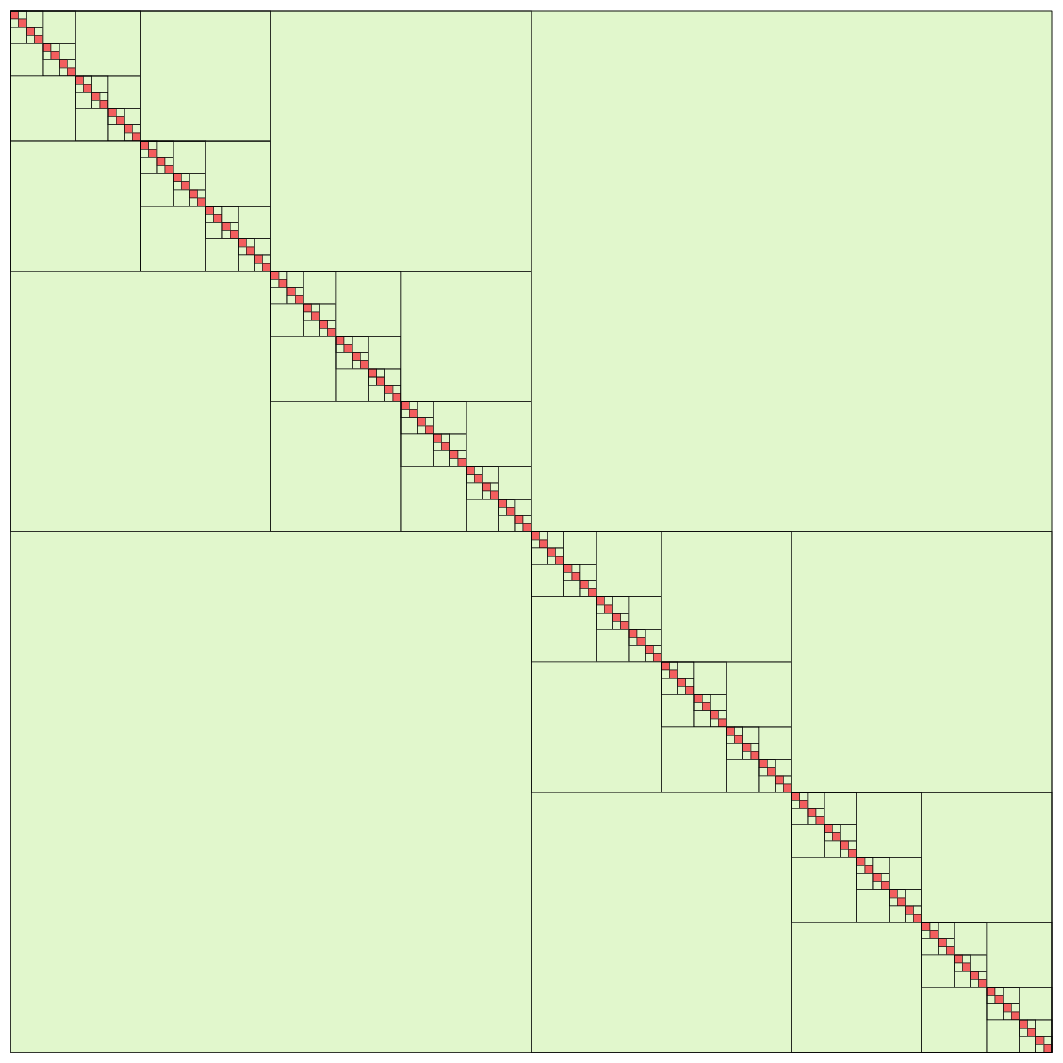}
\caption{Examples of three different block partitioning, generated with three different admissibility criteria: (left) strong, (middle)
  domain-decomposition-based, and (right) weak.}
\label{fig:Hexample_adm}
\end{figure}

Various partitioning strategies for the rows and columns of the matrix and admissibility conditions have been developed
to approximate different types of matrices. Typical admissibility conditions are \emph{strong}, \emph{weak} and based on
domain decomposition \cite{HackHMEng}, for which examples are shown in Figure~\ref{fig:Hexample_adm}. The red blocks
indicate dense or in-admissible blocks whereas green blocks are identified as admissible. The maximal size of the dense
blocks (i.e., how deep the hierarchical subdivision into sub-blocks is) is regulated by the parameter ``$n_{\min}$'', whose
value affects the storage size and the runtime of the \(\H\)-matrix arithmetic, e.g., a smaller value leads to
less storage but is often in-efficient with respect to CPU performance. Typically values of $n_{\min}$ are in the
range $20$ to $150$.


For the computation of the low-rank approximation for admissible sub-blocks many different methods are available, e.g.,
adaptive cross approximation (ACA), hybrid cross approximation (HCA), rank-revealing QR, randomized SVD \cite{TyrtyshACA,ACA2,ACA,Winter,HACA,HonPan:1992,HalMarTro:2011}. For the fixed-rank strategy, the resulting
low-rank matrix is of rank at most \(k\). In case of the fixed-accuracy strategy with a given \(\varepsilon > 0\), the
low-rank approximation $\widetilde \bM$ of the sub-block $\bM$ is computed such that
$\Vert \bM - \widetilde \bM \Vert \le \varepsilon \Vert \bM \Vert $. The storage size of the resulting \mcH-matrix is of order \landau{ k n \log
  n } \cite{GH03}.

\begin{figure}[h!]
\centering
\includegraphics[width=0.33\textwidth]{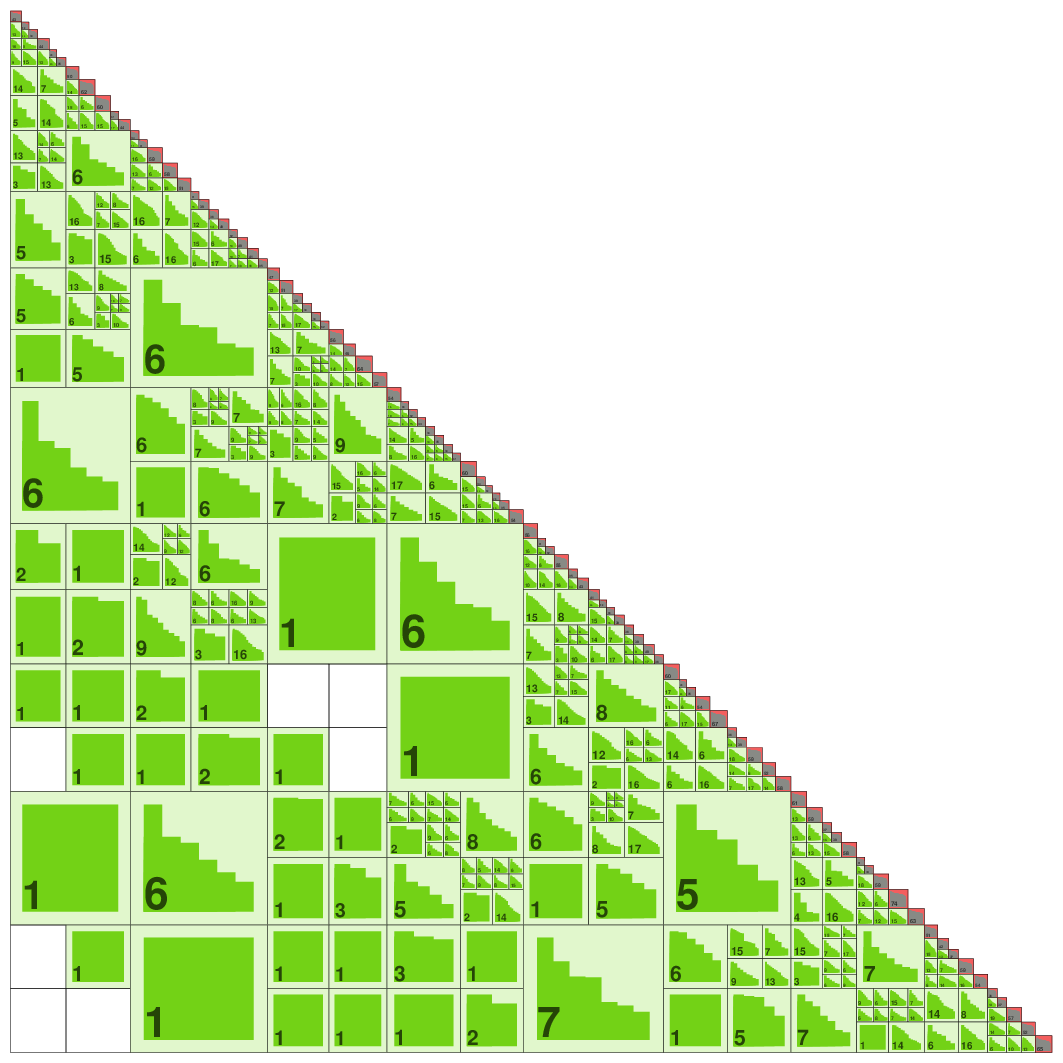}
\includegraphics[width=0.33\textwidth]{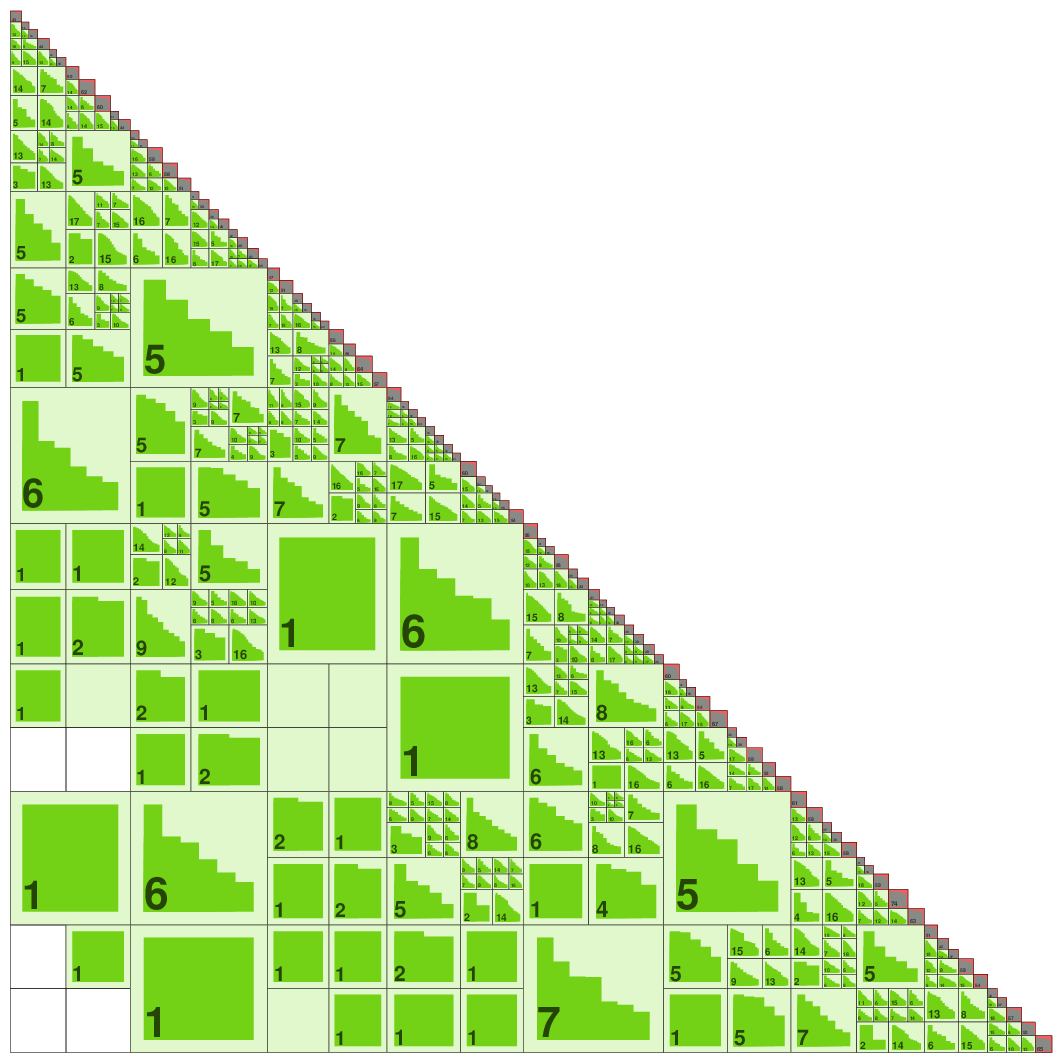}
\includegraphics[width=0.3\textwidth]{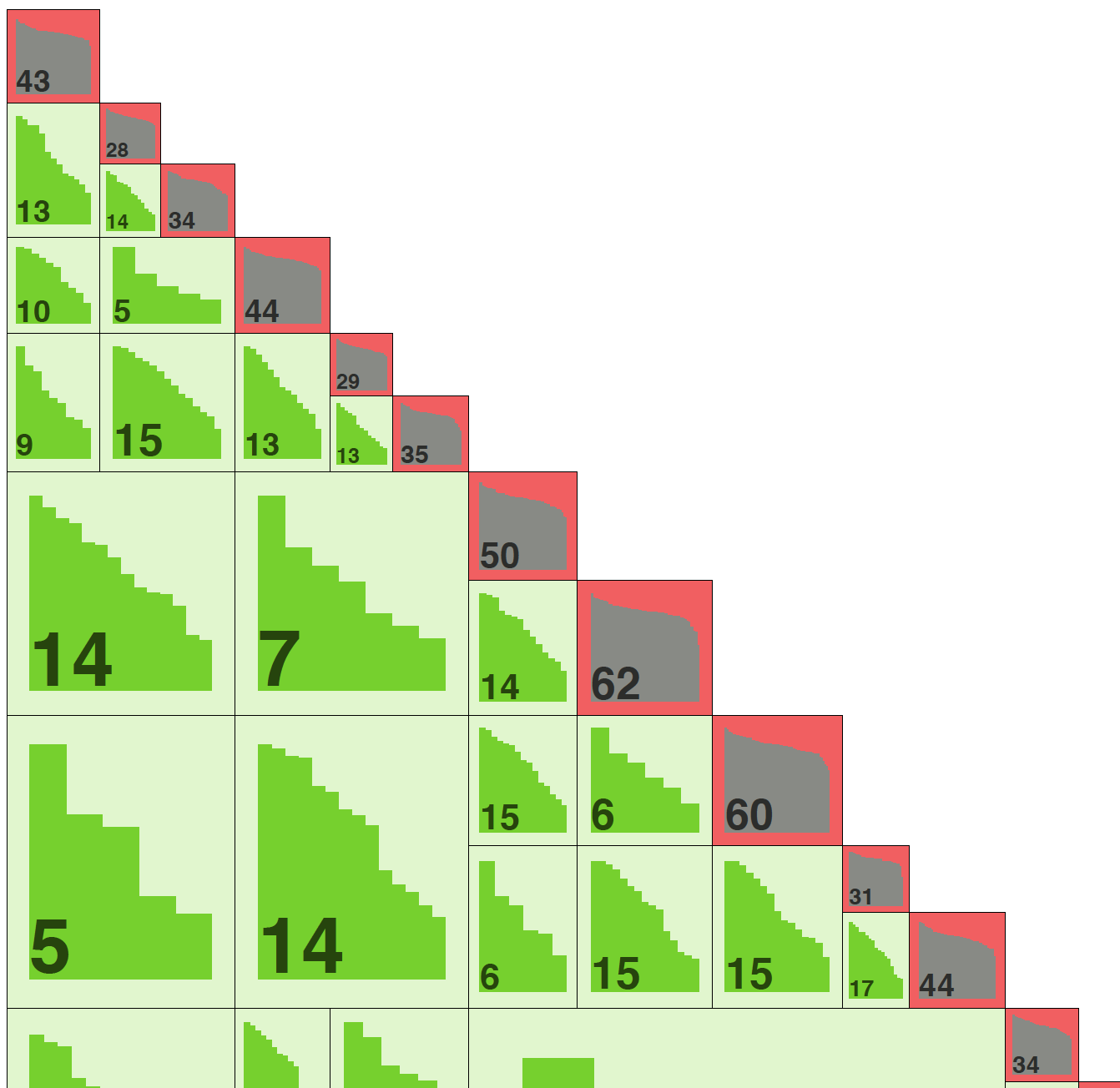}
\caption{Examples of $\H$-matrix approximations of the exponential covariance matrix (left), its hierarchical Cholesky
  factor $\widetilde{\bL}$ (middle), and the zoomed upper-left corner of the matrix (right),  $n=4000$, $\ell=0.09$,
  $\nu=0.5$, $\sigma^2=1$. Approximation and arithmetic performed with a fixed-accuracy of $10^{-5}$. The number inside
  a sub-block indicates the maximal rank, while the ``stairs'' represent its singular values in logarithmic scaling.}
\label{fig:HexampleChol}
\end{figure}

In Figure~\ref{fig:HexampleChol} (left), an example of an \mcH-matrix approximation to \(\bC(\btheta)\) can be found. There,
the local ranks and the decay of singular values in the admissible blocks (green) in logarithmic scale are shown.

In addition to efficient matrix approximation, \mcH-matrices also permit full matrix arithmetic, e.g., matrix addition,
matrix multiplication, inversion or factorization. However, similar to matrix compression, \mcH-matrix arithmetic is
approximate to maintain log-linear complexity. The approximation during arithmetic is again either of a fixed-rank or a
fixed-accuracy \cite{GH03}. In this work, we make use of the \mcH-Cholesky factorization of \(\bC(\btheta)\) (see
Figure~\ref{fig:HexampleChol}).

For \(\bC(\thetab)\), the predefined rank (or accuracy \(\varepsilon\)) defines the accuracy of the \mcH-matrix
approximation, for the initial approximation of \(\bC(\thetab)\) as well as for the Cholesky factorization for 
\(\bC(\thetab)^{-1}\).

\begin{figure}[htbp!]
\centering
\includegraphics[width=0.48\textwidth]{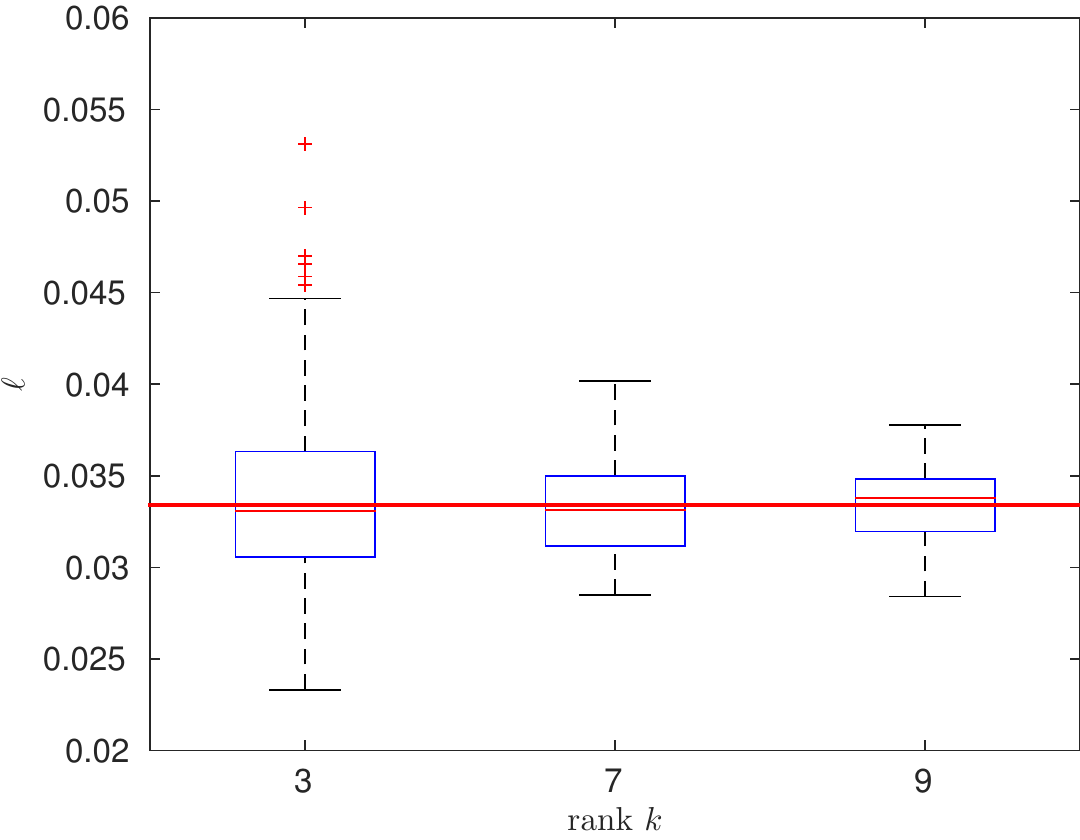}
\includegraphics[width=0.48\textwidth]{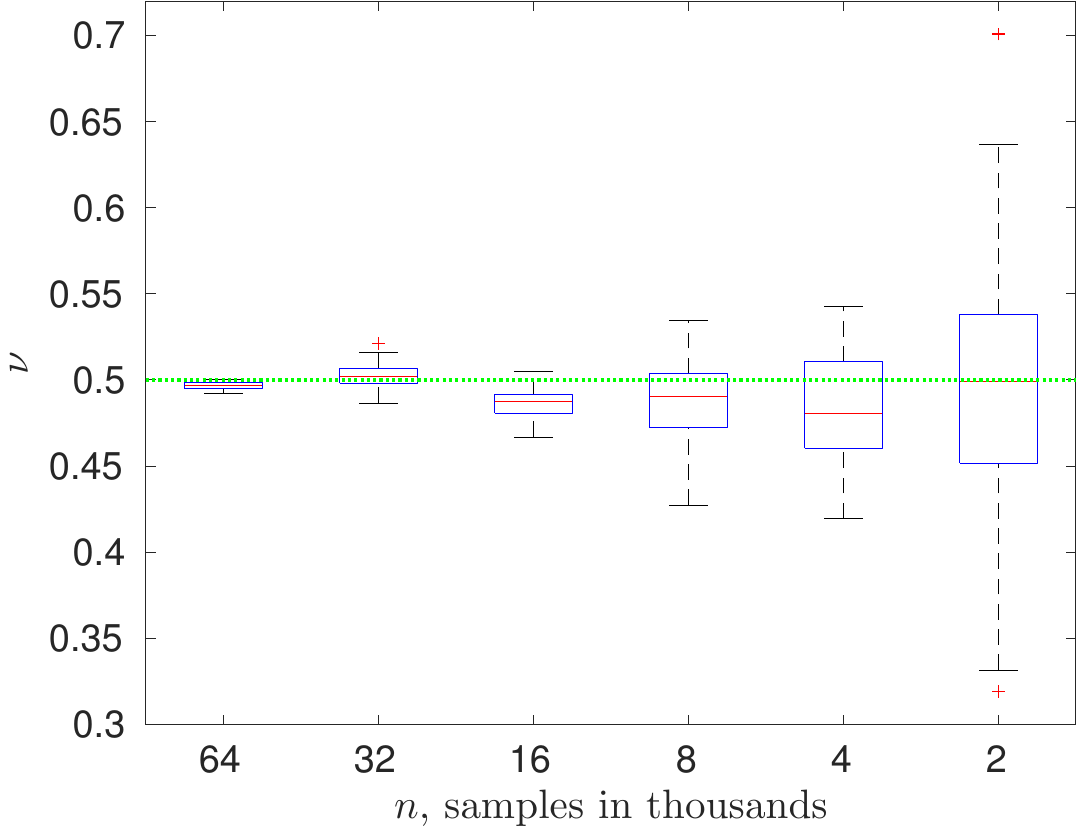}
\caption{(left) Dependence of the boxplots for $\ell$ on the $\H$-matrix rank, when $n=16{,}000$; (right) Convergence of
  the boxplots for $\nu$ with increasing $n$; 100 replicates.} 
\label{fig:Boxranks}
\end{figure}

In Fig.~\ref{fig:Boxranks} (left), the results for computing \(\ell\) with a different rank in the \mcH-matrix
approximation for 100 replicates are shown. On each box, the central red line indicates the median,
the small box indicates the 25$\%$ percentile, and the top (wide) edge indicates the 75$\%$ percentile. The outliers are
marked by the red symbol '+'. The bold long red line denotes the true value of the parameter $\ell=0.0334$. With a larger
rank and hence, with a better approximation, the variance of \(\ell\) decreases.

The dependence of \(\nu\) on the problem size, e.g., the number of measurements is also tested with the results shown in
Figure~\ref{fig:Boxranks} (right). As the results demonstrate, with a larger number the estimation of the parameter
$\nu$ is getting better.
%
%
%

\begin{figure}[htbp!]
\centering
\includegraphics[width=0.48\textwidth]{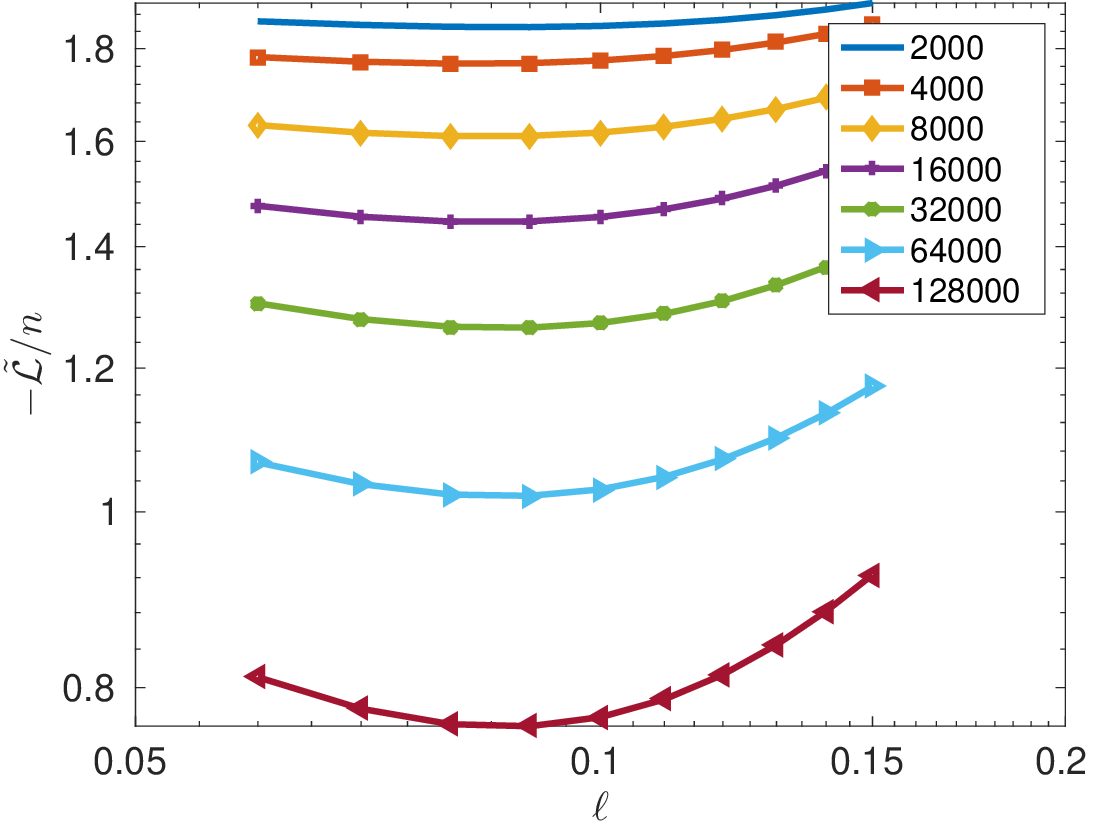}
\includegraphics[width=0.48\textwidth]{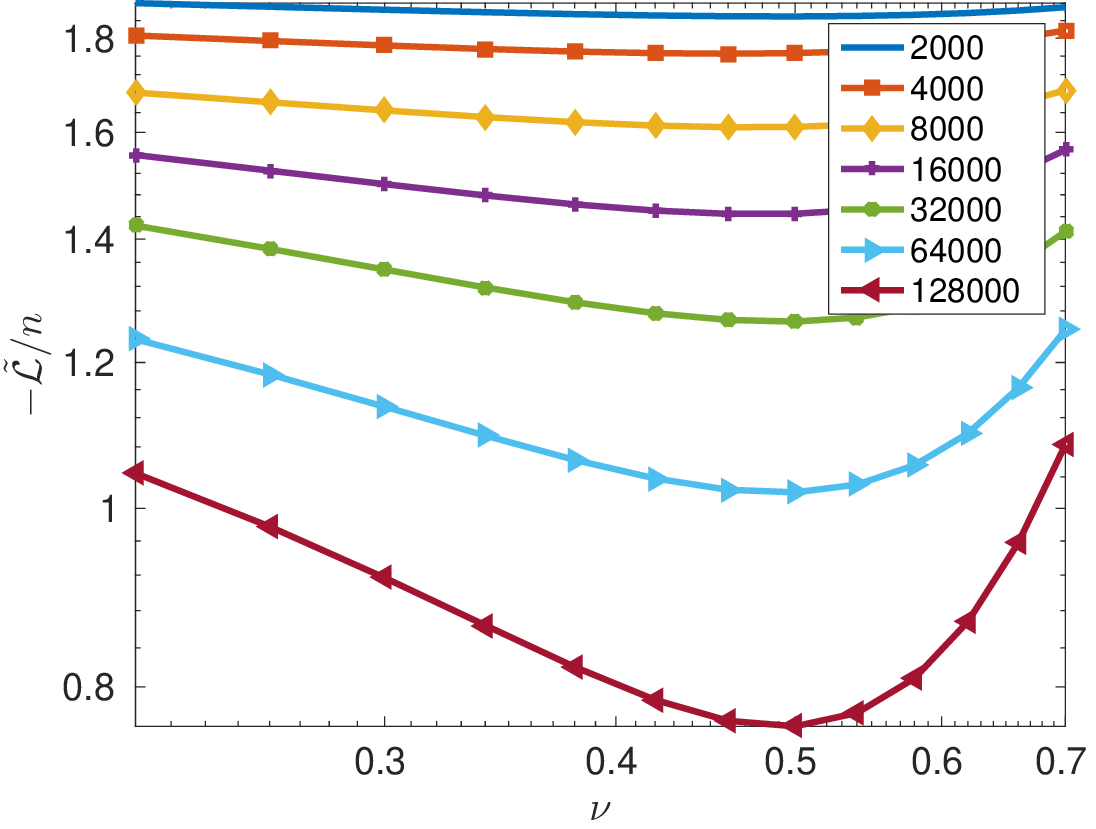}
\caption{(left) Shape of the scaled log-likelihood function, $-\tilde{\LL}/n$, vs. $\ell$ for different sample sizes $n$. (right) Shape of the scaled log-likelihood function, $-\tilde{\LL}/n$, vs. $\nu$ for different sample sizes $n$; }
\label{fig:Shape}
\end{figure}

In Fig.~\ref{fig:Shape}, we illustrate the dependence of $-\tilde{\LL}/n$ on the parameters $\ell$ (left, with
$\nu=0.5$, $\sigma^2=1$), and $\nu$ (right, with $\ell=0.0864$ and $\sigma^2=1$). Both figures demonstrate the smooth
dependance also illustrate the locations of the minima for a different $n$.

%

\subsection{Parallel hierarchical-matrix technique}

We used the parallel $\H$-matrix library HLIBpro \cite{HLIBpro, Ronald14HLU, Ronald08HLU, Ronald08Manual}, which
implements \mcH-matrix approximation and arithmetic functions using a task-based approach to make use of todays
many-core architectures. For this, the mathematical operation is decomposed into small atomic \emph{tasks} with
corresponding incoming and outgoing data dependencies. This set of tasks and dependencies forms a directed acyclic graph
(DAG), which is used for scheduling the tasks to the CPU cores, e.g., if all incoming data dependencies are met, the
corresponding task is executed on the next free CPU core available.

The computational complexity of the different \mcH-matrix operations is shown in Table~\ref{tab:1}. Here, $|V(T)|$
denotes the number of vertices, $|{L}(T)|$ is the number of leaves in the block-cluster tree $T=T_{I\times I}$. The
sequential terms in those estimates are typically due to the sequential behaviour of the corresponding algorithm, e.g.,
strictly following the diagonal during Cholesky factorization, but usually do not show in practical applications since
the majority of the computation work is parallelized.

\begin{table}[h!]
\begin{center}
\caption{Parallel complexity of the main linear operations with further rank truncation in HLIBpro on $p$ cores.}\label{tab:1}
\begin{tabular}{|l|l|} 
\hline 
 Operations &  Parallel Complexity \cite{Ronald05diss} (Shared Memory)\\ 
 with rank truncation & \\ 
 \hline\hline
  build $\widetilde{\bC}$ & $\frac{\mathcal{O}(n\log n)}{p}+\mathcal{O}(\vert V(T) \backslash {L}(T) \vert)$\\ \hline
  store $\widetilde{\bC}$  & $\mathcal{O}(kn\log n)$\\ \hline
  $\widetilde{\bC}\cdot \bz$ & $\frac{\mathcal{O}(kn\log n)}{p}$\\ \hline 
    $\alpha \widetilde{\bA}\oplus \beta \widetilde{\bB}$ & $\frac{\mathcal{O}(n\log n)}{p}$\\ \hline
 $\alpha \widetilde{\bA}\odot \widetilde{\bB}\oplus\beta \widetilde{\bC}$ & $\frac{\mathcal{O}(n\log n)}{p}+\mathcal{O}(\vert V(T)\vert)$\\ \hline
  $\widetilde{\bC}^{-1}$ &  $\frac{\mathcal{O}(n\log n)}{p}+\mathcal{O}(nn_{\min}^2)$\\ \hline
  $\H$-Cholesky $\widetilde \bL$& $\frac{\mathcal{O}(n\log n)}{p}+\mathcal{O}(\frac{k^2n\log^2 n}{n^{1/d}})$, $d=1,2,3$\\ \hline
  determinant $\mydet{\widetilde{\bC}}$  & $\frac{\mathcal{O}(n\log n)}{p}+\mathcal{O}(\frac{k^2n\log^2 n}{n^{1/d}})$, $d=1,2,3$\\ \hline
\end{tabular}
\end{center}
\end{table} 
\section{Memory storage and convergence}
\label{sec:conv}
The Kullback-Leibler divergence (KLD) $D_{KL}(P\Vert Q)$ is a measure of information loss when a distribution $Q$ is used to approximate $P$.  For the multivariate normal distributions $(\bmu_0, \bC)$ and $(\bmu_1, \widetilde{\bC})$, it is defined as follows:
\begin{equation*}
D_{KL}(\bC,\widetilde{\bC})=0.5\left(  
\tr(\widetilde{\bC}^{-1}\bC)+(\bmu_1-\bmu_0)^\top\widetilde{\bC}^{-1}(\bmu_1-\bmu_0)-n-\ln\left (\frac{\mydet{\bC}}{\mydet {\widetilde{\bC}}}\right )
\right ).
\end{equation*}

In Tables~\ref{table:approx_compare05} and \ref{table:approx_compare_rank}, we show the dependence of KLD and two matrix
errors on the $\H$-matrix rank $k$ for the Mat\'{e}rn covariance function with parameters $\ell=\{0.25, 0.75\}$,
$\nu=\{0.5,1.5\}$, and $\sigma^2=\{1.0,1.0\}$, computed on the domain $\mathcal{G}=[0,1]^2$.  All errors are under
control, except for the last column. The ranks $k=10,12$ are too small to approximate the inverse, and, therefore, the
resulting error $\Vert \C (\widetilde{\C})^{-1} -\I \Vert_2 $ is large. Relatively often, the $\H$-matrix procedure, which computes the
$\H$-Cholesky factor $\widetilde{\L}$ or the $\H$-inverse, produces ``NaN'' (not a number) and terminates.  One possible
cause is that some of the diagonal elements can be very close to zero, and their inverse is not defined.  This may
happen when two locations are very close to each other and, as a result, two columns (rows) are linear dependent.  To
avoid such cases, the available data should be preprocessed to remove duplicate locations.  Very often, the nugget
$\tau^2\bI$ is added to the main diagonal to stabilize numerical calculations (see more in Section~\ref{sec:nugget}),
i.e., $\widetilde{\bC}:=\widetilde{\bC}+\tau^2 \bI$. In Tables~\ref{table:approx_compare05} and \ref{table:approx_compare_rank}, the nugget is equal to zero.

\begin{figure}[h!]
\centering
\includegraphics[width=.48\textwidth]{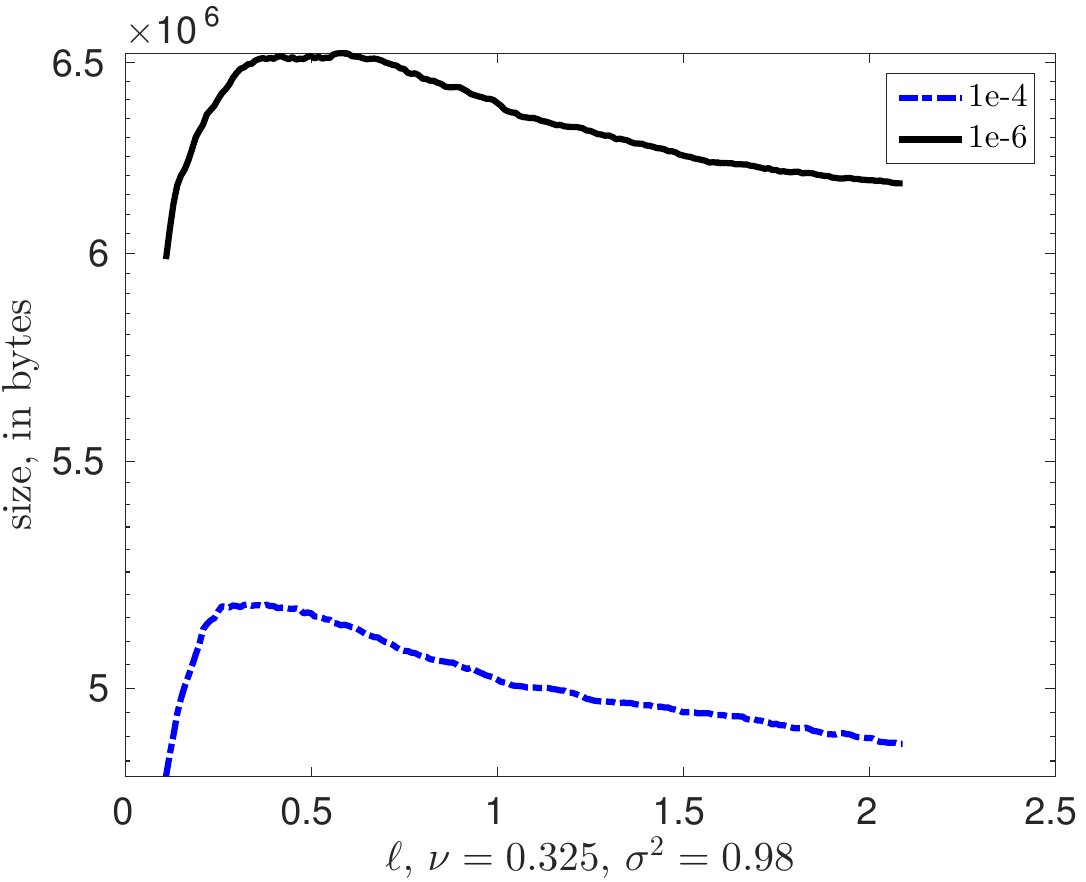}
\includegraphics[width=.48\textwidth]{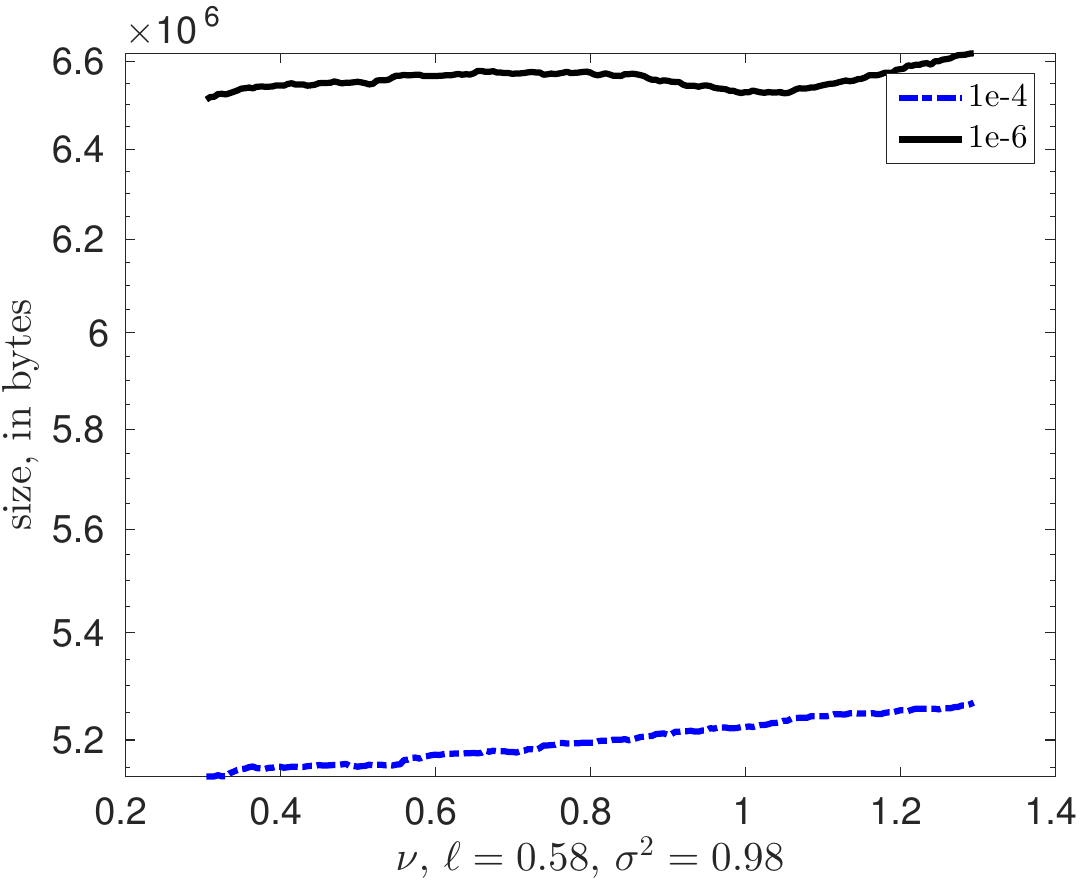}
\caption{(left) Dependence of the matrix size on (left) the covariance length $\ell$, and (right) the smoothness $\nu$ for two different accuracies in the $\H$-matrix sub-blocks $\varepsilon =\{10^{-4}, 10^{-6}\}$, for $n=2,000$ locations in the domain $[32.4,  43.4]\times [-84.8, -72.9]$. }
\label{fig:Stab_size}
\end{figure}

\begin{table}[h!]
\centering
\caption{KLD and $\H$-matrix approximation errors vs. the $\H$-matrix rank $k$ for Mat\'{e}rn covariance function, $\ell=\{0.25, 0.75\}$, $\nu=0.5$, $\sigma^2=1$, domain $\mathcal{G}=[0,1]^2$, and $\Vert C_{(\ell=0.25,0.75)}\Vert_2=\{212, 568\}$.}
\begin{tabular}{|l|ll|ll|ll|}
\hline
  $k$
  & \multicolumn{2}{c|}{KLD}
  & \multicolumn{2}{c|}{$\Vert \C - \widetilde{\C} \Vert_2$}
  & \multicolumn{2}{c|}{$\Vert \C (\widetilde{\C})^{-1} -\bI \Vert_2  $} \\
  &  $\ell=0.25$   &  $\ell=0.75$ &  $\ell=0.25$   &  $\ell=0.75$ &  $\ell=0.25$   &  $\ell=0.75$ \\ 
\hline
 10  &  $2.6\cdot 10^{-3}$ & $2.0\cdot 10^{-1}$   & $7.7\cdot 10^{-4}$ & $7.0\cdot 10^{-4}$ & $6.0\cdot 10^{-2}$ & \(3.1\cdot 10^{0}\)\\ 
 12  &  $5.0\cdot 10^{-4}$ & $2.2\cdot 10^{-2}$   & $9.7\cdot 10^{-5}$ & $5.6\cdot 10^{-5}$ & $1.6\cdot 10^{-2}$ &\(5.0\cdot 10^{-1}\)\\ 
 15  &  $1.0\cdot 10^{-5}$ & $9.0\cdot 10^{-4}$   & $2.0\cdot 10^{-5}$ & $1.1\cdot 10^{-5}$ & $8.0\cdot 10^{-4}$ & $2.0\cdot 10^{-2}$\\ 
 20  &  $4.5\cdot 10^{-7}$ & $4.8\cdot 10^{-5}$   & $6.5\cdot 10^{-7}$ & $2.8\cdot 10^{-7}$ & $2.1\cdot 10^{-5}$ & $1.2\cdot 10^{-3}$\\ 
 50  &  $3.4\cdot 10^{-13}$& $5.0\cdot 10^{-12}$  & $2.0\cdot 10^{-13}$& $2.4\cdot 10^{-13}$& $4.0\cdot 10^{-11}$  & $2.7\cdot 10^{-9}$\\ 
\hline
\end{tabular}
\label{table:approx_compare05}  
\end{table}
\begin{table}[h!]
\centering
\caption{KLD and $\H$-matrix approximation error vs. the $\H$-matrix rank $k$ for Mat\'{e}rn covariance function, $\ell=\{0.25, 0.75\}$, $\nu=1.5$, $\sigma^2=1$, domain $\mathcal{G}=[0,1]^2$, and $\Vert \C_{(\ell=0.25,0.75)}\Vert_2=\{720, 1068\}$.}
\begin{tabular}{|l|ll|ll|ll|}
\hline
 $k$ & \multicolumn{2}{c|}{KLD}& \multicolumn{2}{c|}{$\Vert \C - \widetilde{\C} \Vert_2$} & \multicolumn{2}{c|}{$\Vert \C (\widetilde{\C})^{-1} -\bI \Vert_2  $} \\
     &  $\ell=0.25$   &  $\ell=0.75$ &  $\ell=0.25$   &  $\ell=0.75$ &  $\ell=0.25$   &  $\ell=0.75$ \\ 
\hline
 20  &  $1.2\cdot 10^{-1}$& $2.7\cdot 10^{0}$ & $5.3\cdot 10^{-7}$& $2.3\cdot 10^{-7}$   & $4.5\cdot 10^{0}$& $7.2\cdot 10^{1}$\\ 
 30  &  $3.2\cdot 10^{-5}$& $4.0\cdot 10^{-1}$  & $1.3\cdot 10^{-9}$& $5.0\cdot 10^{-10}$  & $4.8\cdot 10^{-3}$& $2.0\cdot 10^{1}$\\ 
 40  &  $6.5\cdot 10^{-8}$& $1.0\cdot 10^{-2}$  &$1.5\cdot 10^{-11}$& $8.0\cdot 10^{-12}$  & $7.4\cdot 10^{-6}$& $5.0\cdot 10^{-1}$\\ 
 50  &  $8.3\cdot 10^{-10}$& $3.0\cdot 10^{-3}$ &$2.0\cdot 10^{-13}$& $1.5\cdot 10^{-13}$&  $1.5\cdot 10^{-7}$& $1.0\cdot 10^{-1}$\\ 
\hline
\end{tabular}
\label{table:approx_compare_rank}  
\end{table}
\newpage
Figure~\ref{fig:Stab_size} shows that the $\H$-matrix storage cost 
remains almost the same for the different parameters $\ell=\{0.15,...,2.2\}$ (left) and $\nu=\{0.3,...,1.3\}$ (right). The computational domain is  $[32.4,  43.4]\times [-84.8, -72.9]$ with $n=2{,}000$.

\begin{figure}[h!]
\centering
\includegraphics[width=.48\textwidth]{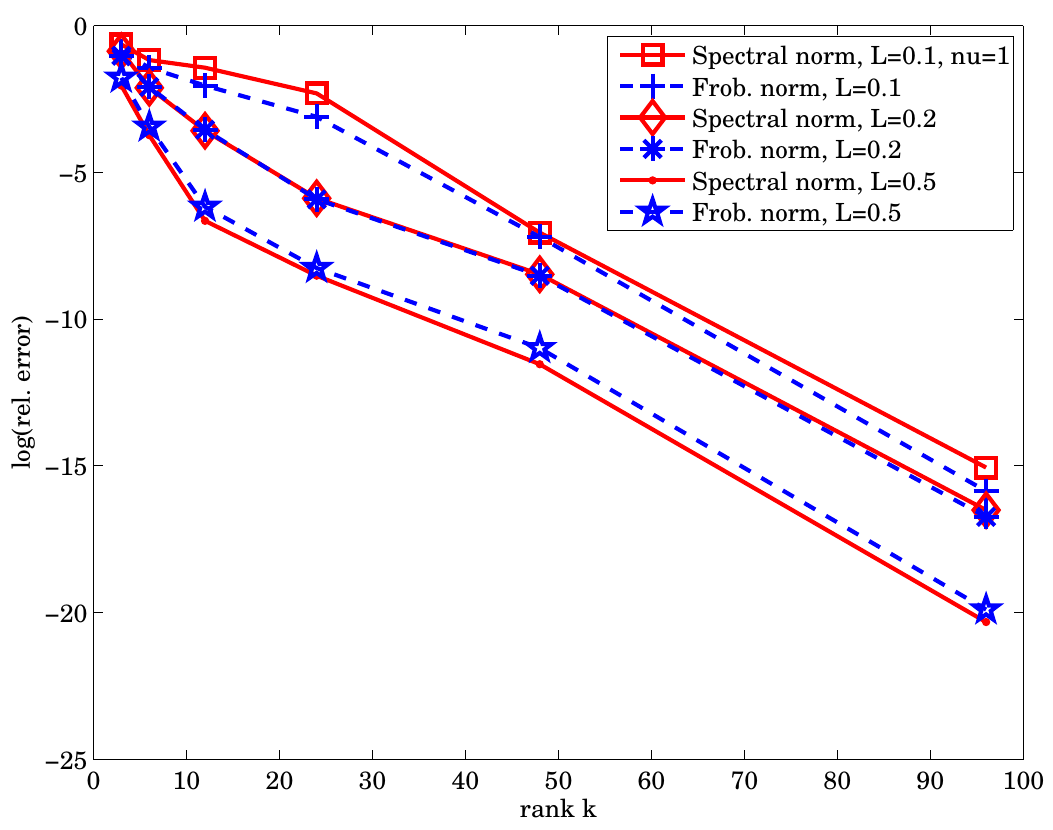}
\includegraphics[width=.48\textwidth]{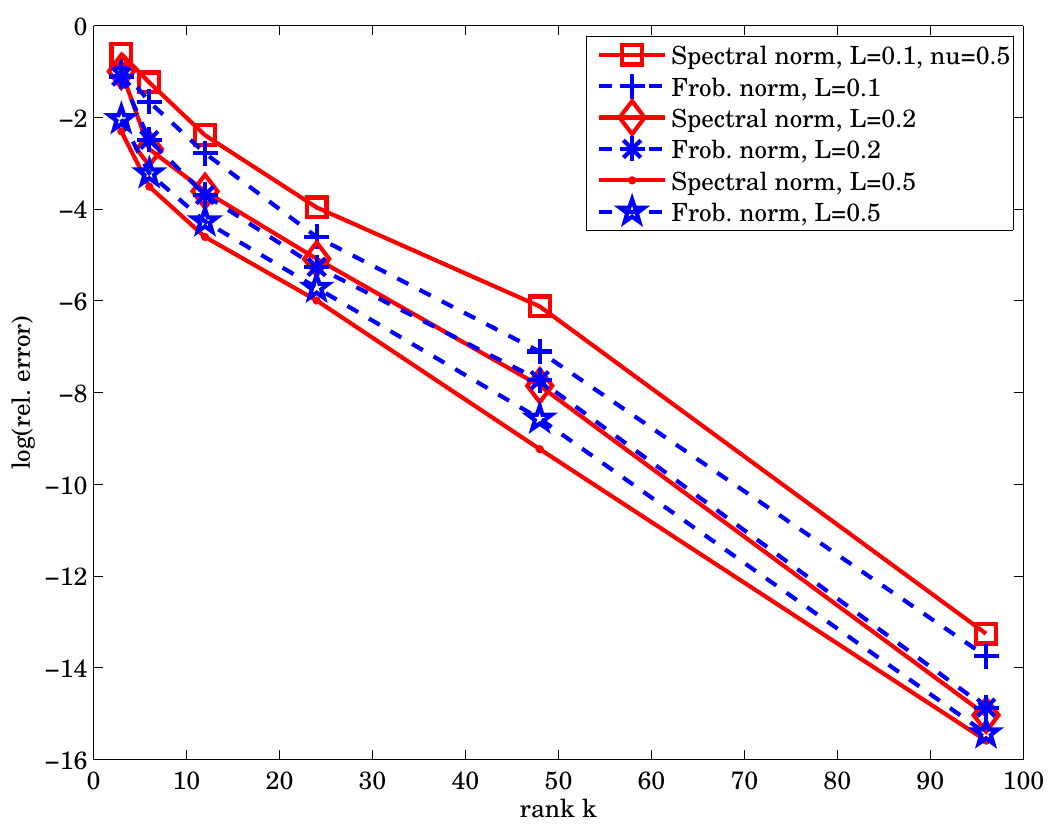}
\caption{Convergence of the $\H$-matrix approximation errors for covariance lengths $\{0.1,0.2, 0.5\}$; (left) $\nu=1$ and (right) $\nu=0.5$, computational domain $[0,1]^2$.}
\label{fig:Convnu}
\end{figure}
\begin{figure}[h!]
\centering
\includegraphics[width=.48\textwidth]{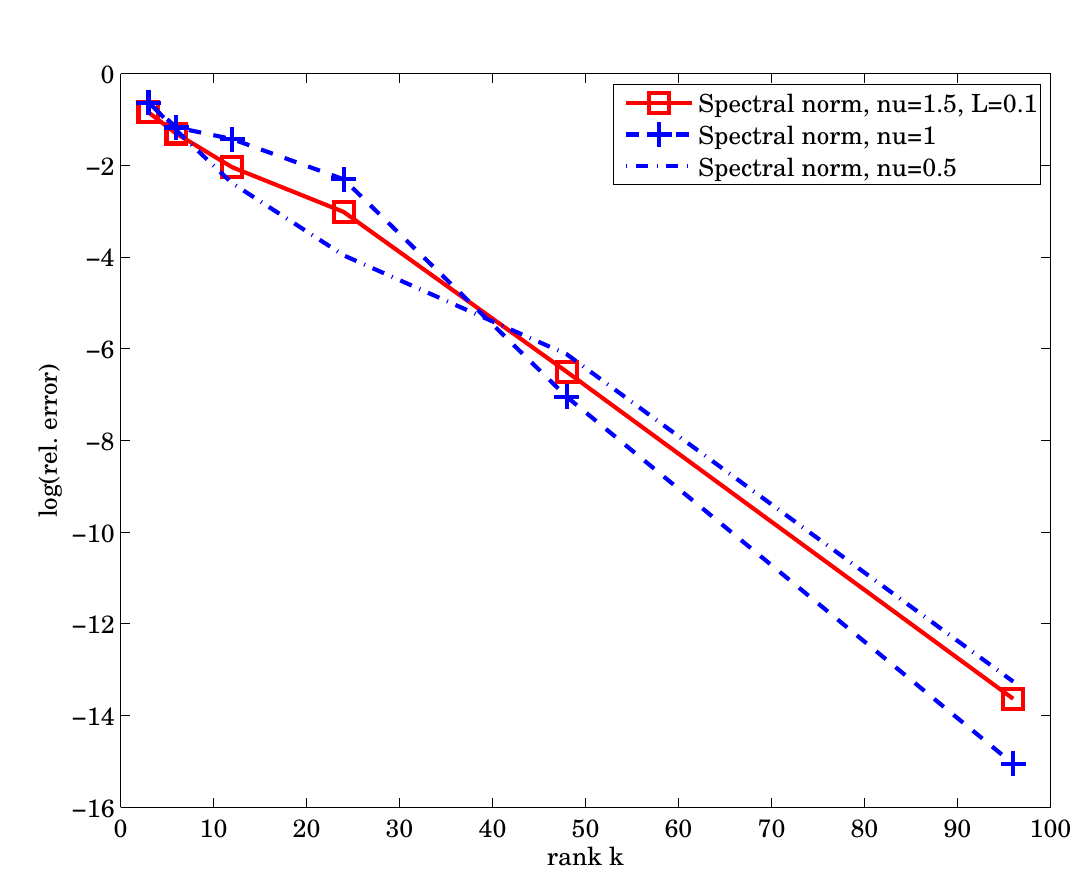}
\includegraphics[width=.48\textwidth]{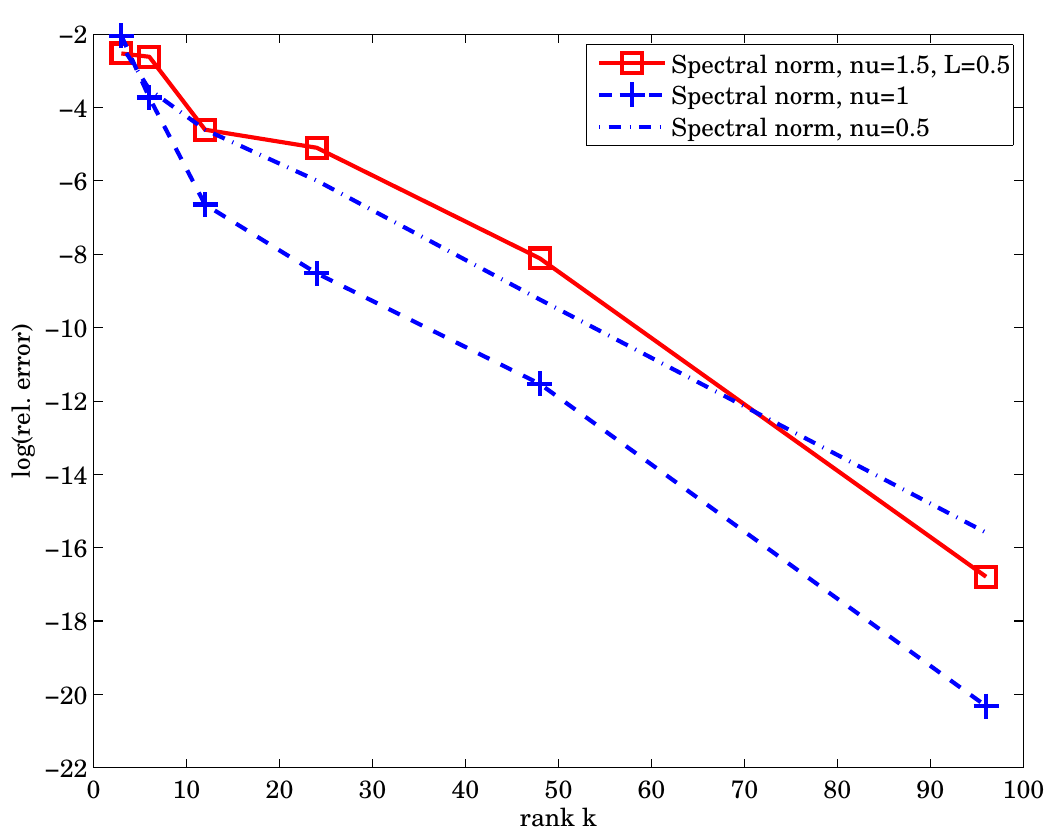}
\caption{Convergence of the $\H$-matrix approximation errors for $\nu=\{0.5,1,1.5\}$; (left) covariance length $0.1$  and (right) covariance length $0.5$, computational domain $[0,1]^2$.}
\label{fig:Conv}
\end{figure}

In Figure~\ref{fig:Convnu}, we plot the convergence of $\Vert \bC - \widetilde{\bC}\Vert$ in the Frobenius and spectral norms vs. the rank $k$ for different covariance lengths. The smoothness parameter is equal to 1 (left), and 0.5 (right). In Figure~\ref{fig:Conv}, we plot $\Vert \bC - \widetilde{\bC}\Vert_2$ vs. the rank $k$ for different smoothness parameters. The covariance length is equal to 0.1(left), and 0.5 (right). The computational domain in both cases was a unit square $[0,1]^2$.

%
%
%
\section{Software installation}
\label{sec:install}
This section contains a summary of the information provided at
\url{https://www.hlibpro.com} and 
\url{https://github.com/litvinen/HLIBCov.git}\ .
HLIBpro supports both shared and distributed memory architectures, though in this work we only use the shared memory
version. For the implementation of the task-parallel approach, Intel's Threading Building Blocks (TBB) is used. HLIBpro
is free for academic purposes, and is distributed in a pre-compiled form (no source code available). 
Originally, HLIBpro was developed for solving FEM and BEM problems \cite{Ronald08HLU, Ronald14HLU}.
In this work, we extend the applicability of HLIBpro to dense covariance matrices and log-likelihood functions.
\paragraph{Installation:} HLIBCov uses the functionality of HLIBpro; therefore, HLIBpro must be installed first. All
functionality implemented by HLIBCov is based on HLIBpro, i.e., no extra software is needed in addition to the libraries
needed by HLIBpro. This also holds for the Mat\'ern kernel, which uses Bessel functions and maximization algorithms,
both being provided by the GNU Scientific Library (GSL) and also used by HLIBpro. The reader can easily replace GSL with
his own optimization library. The Bessel functions are also available in other packages. 

To install HLIBpro on MacOS and Windows, we refer the reader to \url{www.HLIBpro.com} for further details.

\begin{table}[h!]
  \caption{Version of Software used for Experiments}\label{tab:software}
  \centering
  \begin{tabular}{ll}
    \multicolumn{1}{c}{\textbf{Software}} & \multicolumn{1}{c}{\textbf{Version}} \\
    \hline
    HLIBCov & 1.0 \\
    HLIBpro & 2.6 \\
    GSL & 1.16 \\
    TBB & 4.3
  \end{tabular}
\end{table}
\paragraph{Hardware.} All of the numerical experiments herein are performed on a Dell workstation with two Intel(R)
Xeon(R) E5-2680 v2 CPUs (2.80GHz, 10 cores/20 threads) and 128 GB main memory.
\paragraph{Adding HLIBCov to HLIBpro.}
The easiest form of compiling HLIBCov is by using the compilation system of HLIBpro. For this, the source code file of
HLIBCov is placed in the \textit{examples} directory of HLIBpro and an entry is added to the file \textit{examples/SConscript}:
\begin{lstlisting}
  $ examples.append(cxxenv.Program('loglikelihood.cc'))
\end{lstlisting}
Afterwards, the make process of HLIBpro is run to compile also HLIBCov (see HLIBpro installation instructions at
\url{www.hlibpro.com}).
%
\paragraph{Input of HLIBCov.}
The input contained in the first line is the total number of locations $N$.
Lines $2,...,N+1$ contain the coordinates $x_i$, $y_i$, and the measurement value. An example is provided below;
\begin{lstlisting}
3
0.1  0.2  88.1
0.1  0.3  87.2
0.2  0.4  86.0
\end{lstlisting}
HLIBpro requires neither a list of finite elements nor a list of edges. We provide several examples of few input files of different size on the open-access file hosting service GitHub
(\url{https://github.com/litvinen/HLIBCov.git}).
We added two data sets to GitHub: data.tar.gz and moisture$\_$data.zip. Both examples contain multiple data sets of different sizes.
%
\paragraph{Output of HLIBCov.}
The main output is the three identified parameter values $\thetab=(\ell,\nu,\sigma^2)^\top$. The auxiliary output may include many details: $\H$-matrix details (the maximal rank $k$, the maximal accuracy in each sub-block, and the Frobenius and spectral norms of $\widetilde{\bC}$, $\widetilde{\bL}$, ${\widetilde{\bL}}^{-1}$, $\Vert \bI-{\widetilde{\bL}\widetilde{\bL}^\top}^{-1} \Vert$). Additionally, iterations of the maximization algorithm can also be printed out. The example of an output file 
provided below contains two iterations: the index, $\nu$, $\ell$, $\sigma^2$, $\widetilde{\LL}$, and the residual $\mbox{TOL}$ of the iterative method:
\begin{lstlisting}
1 0.27    2.4   1.30  L = 1762.1  TOL= 0.007
2 0.276  2.41 1.29  L = 1757.2  TOL= 0.009
\end{lstlisting}
If the iterative process is converging, then the last row will contain the solution $\thetab^{*}=(\ell^{*},\nu^{*},{\sigma^{*}}^2)^\top$.
When computing error boxes, the output file will contain $M$ solutions ($n$, $\ell^{*}$, $\nu^{*}$, ${\sigma^{*}}^2$), where $M$ is the number of replicates:
\begin{lstlisting}
4000 0.54  0.082  1.01
4000 0.53  0.083  1.02
4000 0.55  0.081  1.02
\end{lstlisting}
The name of the output file can be found in the main() procedure in loglikelihood.cc.
\section{Numerical experiments}
\label{sec:MC}
We generate a sample set with parameters $(\ell^*, \nu^*,{\sigma^*}^2)=(0.0864, 0.5, 1.0)$ and then try to infer these parameters.
%
%
\subsection{Generation of the synthetic data}
To build $M$ various data-sets ($M$ replicates) with $n\in \{ 64,...,4,2\}\times 1000 $ locations, we generate a large vector $\bZ_0$ with $n_0=2\cdot 10^6$ locations, and randomly sample $n$ points from it. We note that if the locations are very close to each other, then the covariance matrix may be singular or the Cholesky factorization will be very difficult to compute. 

\begin{figure}[h!]
\centering
\includegraphics[width=0.49\textwidth]{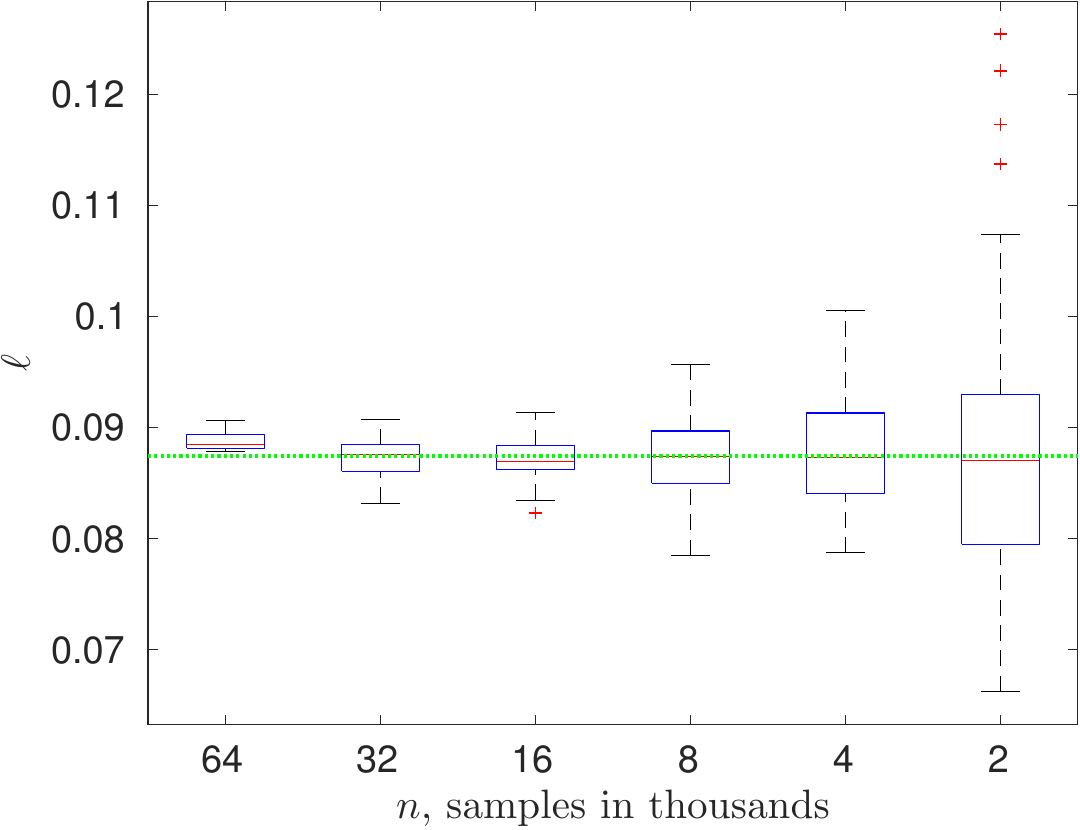}
\includegraphics[width=0.49\textwidth]{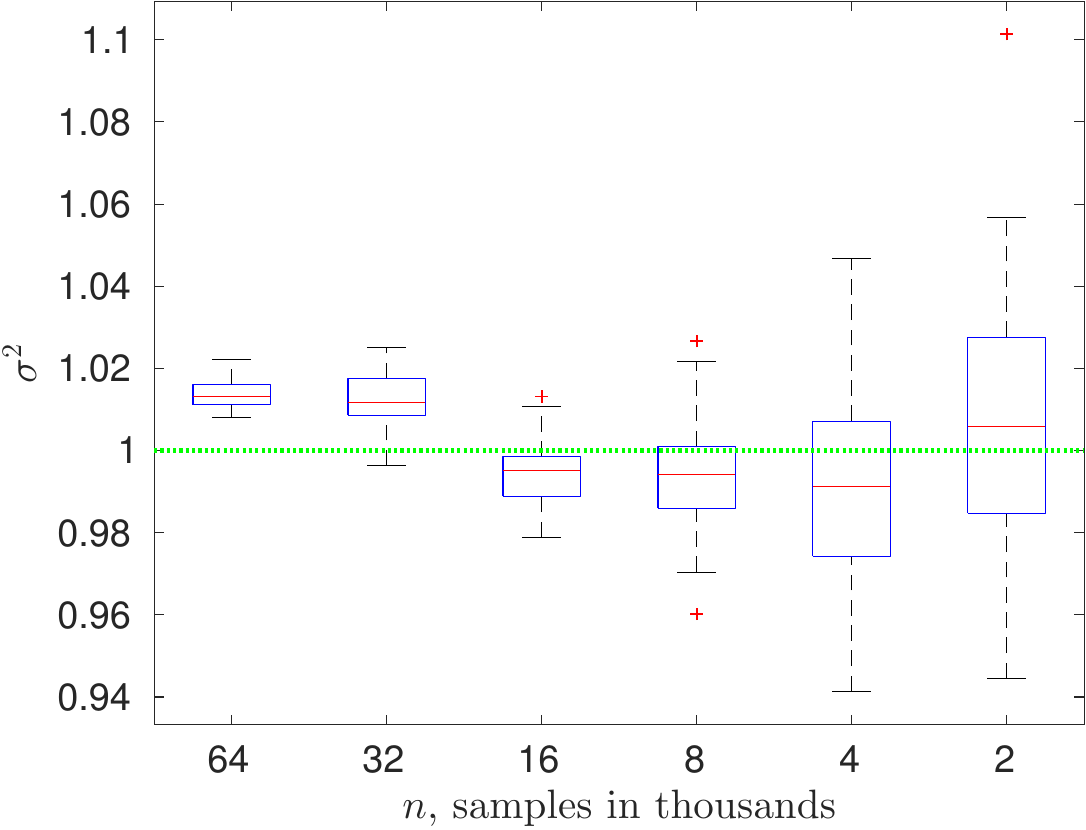}
\caption{Synthetic data with known parameters $(\ell^*, \nu^*, {\sigma^*}^2)=(0.0864,0.5,1.0)$. Boxplots for $\ell$ and $\sigma^2$ for $n=1,000\times \{64,32,...,4,2\}$; 100 replicates.}
\label{fig:BoxPlot}
\end{figure}

To generate the random data $\bZ_0\in \mathbb{R}^{n_0}$, we compute the $\H$-Cholesky factorization of $\bC(0.086, 0.5, 1.0)= \widetilde{\bL}\widetilde{\bL}^\top$. Then, we evaluate $\bZ_0=\widetilde{\bL} \xib$, where $\xib\in \mathbb{R}^{n_0}$ is a normal vector with zero mean and unit variance. We generate $\bZ_0$ only once. 
Next, we run our optimization algorithm and try to identify (recover) the ``unknown" parameters $(\ell, \nu,\sigma^2)^\top$. The resulting boxplots for $\ell$ and $\sigma^2$ over $M=100$ replicates are illustrated in Fig.~\ref{fig:BoxPlot}. We see that the variance (or uncertainty) decreases with increasing $n$. The green line indicates the true values.

%
To identify all three parameters simultaneously, we solve a three-dimensional optimization problem. The maximal number of iterations is set to 200, and the residual is $10^{-6}$.
The behavior and accuracy of the boxplots depend on the $\H$-matrix rank, the maximum number of iterations to achieve a certain threshold, the threshold (or residual) itself, the initial guess, the step size in each parameter of the maximization algorithm, and the maximization algorithm. All replicates of $\bZ$ are sampled from the same generated vector of size $n_0=2\cdot 10^6$. 

In Table~\ref{table:approx_compare15}, we present the almost-linear storage cost (columns 3 and 6) and the computing time (columns 2 and 5).\\ 
\begin{table}[h!]
\centering
\caption{Computing time and storage vs $n$. The number of parallel computing cores is 40, $\hat{\nu}=0.33$, $\hat{\ell}=0.65$, $\hat{\sigma^2}=1.0$. $\H$-matrix accuracy in each sub-block for both $\widetilde{\bC}$ and $\widetilde{\bL}$ is $10^{-5}$.}
\begin{small}
\begin{tabular}{|c|ccc|ccc|}
\hline
$ n$ & \multicolumn{3}{c|}{$\widetilde{\bC}$} &  \multicolumn{3}{c|}{$\widetilde{\bL}\widetilde{ \bL}^\top$}  \\
             & comp. time & size & kB/dof  & comp. time & size &  $\Vert \bI-(\widetilde{\bL}\widetilde{\bL}^\top)^{-1}\widetilde{\bC} \Vert_2$     \\ 
        &  sec. & MB &     & sec. &  MB &      \\ 
\hline
 32{,}000        & 3.3 & 162   & 5.1 & 2.4 & 172.7 & $2.4\cdot 10^{-3}$  \\ 
128{,}000      & 13.3 & 776   & 6.1 & 13.9 & 881.2 & $1.1\cdot 10^{-2}$  \\ 
512{,}000      & 52.8 & 3420   & 6.7 & 77.6 & 4150 & $3.5\cdot 10^{-2}$  \\ 
2{,}000{,}000 & 229 & 14790   & 7.4 & 473 & 18970 & $1.4\cdot 10^{-1}$   \\ 
\hline
\end{tabular}
\end{small}
\label{table:approx_compare15}  
\end{table}
\begin{figure}[htbp!]
\centering
\includegraphics[width=0.3\textwidth]{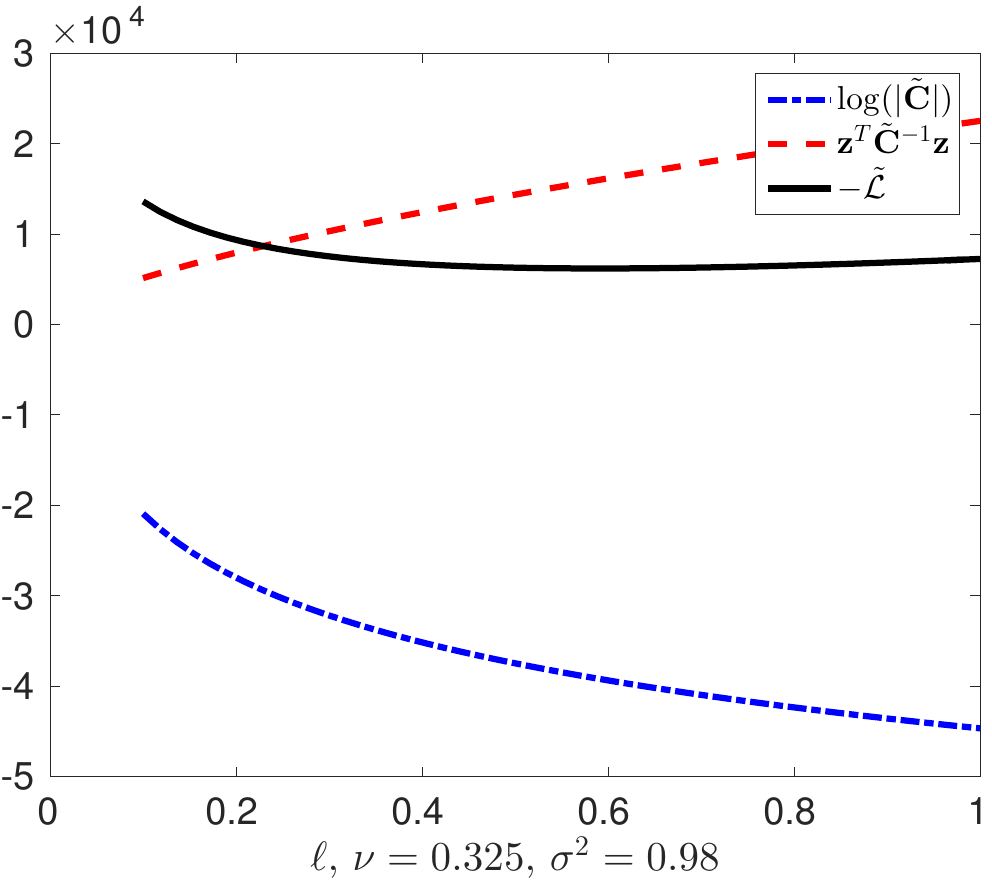}
\includegraphics[width=0.3\textwidth]{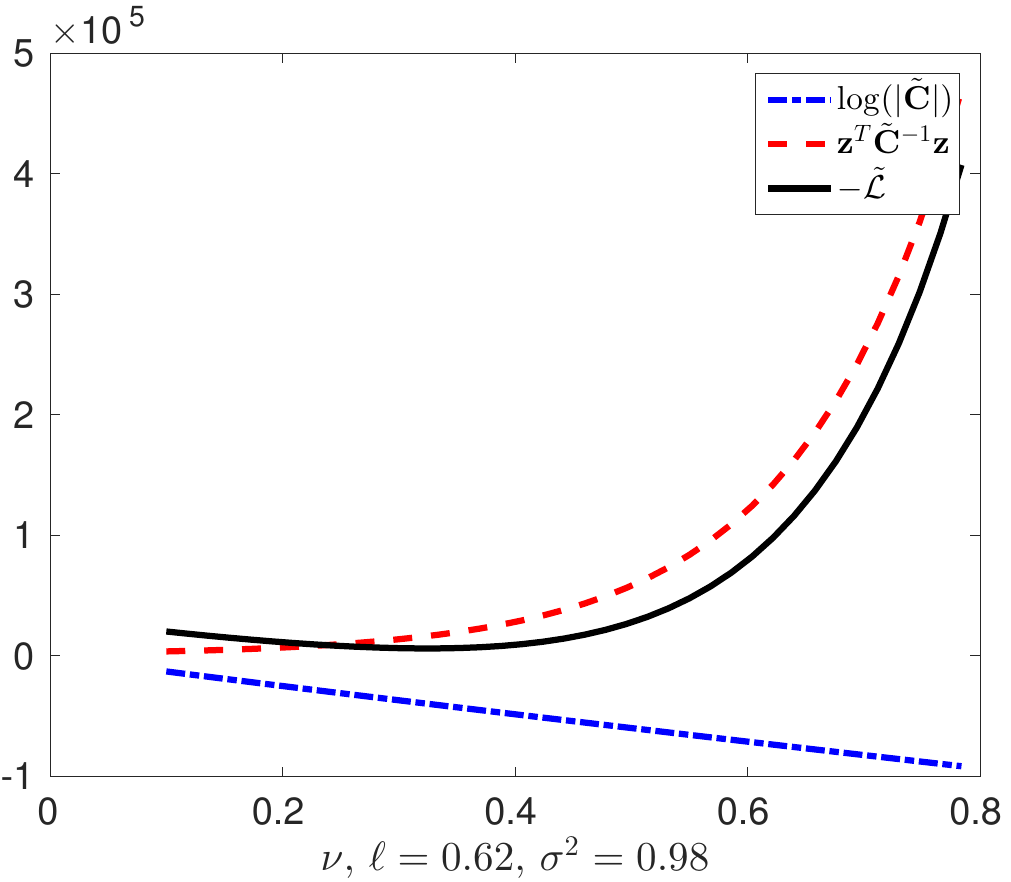}
\includegraphics[width=0.3\textwidth]{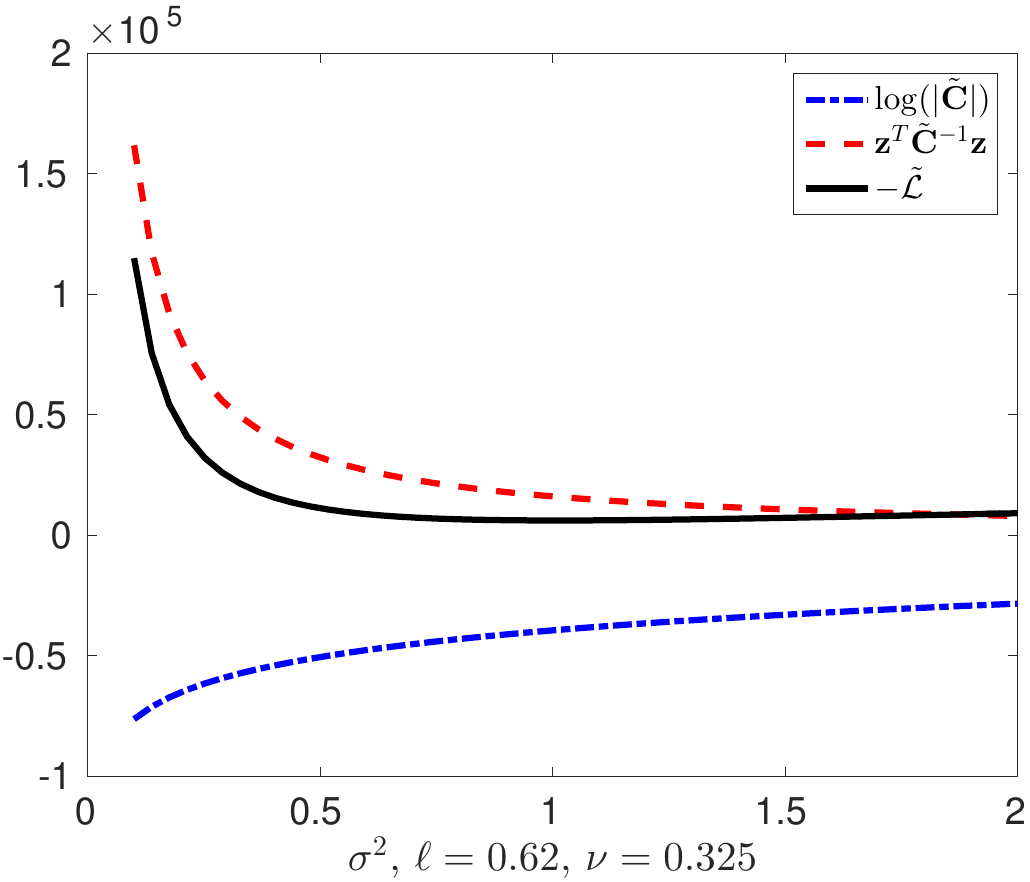}
\caption{Dependence of the negative log-likelihood and its ingredients on parameters $\ell$ (on the left); $\nu$ (in the middle); and $\sigma^2$ (on the right). In each experiment the other two parameters are always fixed. $n=64,000$.}
\label{fig:all_shapes}
\end{figure}

The shape of the negative log-likelihood function and its components are illustrated in Fig.~\ref{fig:all_shapes}. This helps us to understand the behavior of the iterative optimization method, and the contributions of the log-determinant and the quadratic functional. We see that the log-likelihood is almost flat, and that it may be necessary to perform many iterations in order to find the minimum.

\begin{table}[h!]
\centering
\caption{Comparison of three log-likelihood functions computed with three different $\H$-matrix accuracies $\{10^{-7},10^{-9},10^{-11}\}$. Exponential covariance function discretized in the domain $[32.4,  43.4]\times [-84.8, -72.9]$, $n=32{,}000$ locations. Columns correspond to different covariance lengths $\{0.001,...,0.1\}.$}
\begin{tabular}{|c|c|c|c|c|c|c|c|c|}
\hline
 $\ell$                 &  0.001      &  0.005    & 0.01      &  0.02  & 0.03      & 0.05       & 0.07      &0.1  \\  \hline
 $-\widetilde{\LL}(\ell;10^{-7})$&  44657  & 36157  &36427 & 40522 & 45398 &  68450 & 70467 & 90649\\ \hline
 $-\widetilde{\LL}(\ell;10^{-9})$&  44585  & 36352  &36113 & 41748 & 47443 &  60286 & 70688 & 90615\\ \hline
$-\widetilde{\LL}(\ell;10^{-11})$& 44529  & 37655  &36390 & 42020 & 47954 &  60371 & 72785 & 90639\\ \hline
\end{tabular}
\label{table:covfunctions}  
\end{table}
\subsection{Adding nugget $\tau^2$} 
\label{sec:nugget}
When the diagonal values of $\widetilde{\bC}$ are very close to zero, $\H$-Cholesky becomes unstable producing negative
entries on the diagonal during computation. By adding a diagonal matrix with small positive numbers, all the singular
values are increased and moved away from zero. However, by adding a nugget, we redefine the original matrix as
$\widetilde{\bC}:=\widetilde{\bC}+\tau^2 \bI$. Below, we analyze how the loglikelihood function, as well as its maximum
are changing by this.

We assume $\mydet{\widetilde{\bC}}\neq 0$. For a small perturbation matrix $\bE$ \cite{DemmelErr92}, it holds that
\begin{equation*}
\frac{\Vert (\widetilde{\bC}+\bE)^{-1} - (\widetilde{\bC})^{-1} \Vert}{\Vert \widetilde{\bC}^{-1} \Vert}\leq\kappa(\bC)\cdot \frac{\Vert \bE \Vert}{\Vert \widetilde{\bC} \Vert}= \frac{\kappa(\widetilde{\bC}) \tau^2 }{\Vert {\widetilde{\bC}} \Vert},
\end{equation*}
where $\kappa(\widetilde{\bC})$ is the condition number of $\widetilde{\bC}$, and $\bE=\tau^2\bI$.
Alternatively, by substituting $\kappa(\widetilde{\bC}):=\Vert \widetilde{\bC} \Vert \cdot \Vert \widetilde{\bC}^{-1} \Vert$, we obtain
\begin{equation}
\label{eq:relerror}
\frac{\Vert (\widetilde{\bC}+\tau^2 \bI)^{-1} - (\widetilde{\bC})^{-1} \Vert}{\Vert \widetilde{\bC}^{-1} \Vert}\leq  \tau^2 \Vert \widetilde{\bC}^{-1} \Vert.
\end{equation}
From \eqref{eq:relerror}, we see that the relative error on the left-hand side of \eqref{eq:relerror} depends on the norm $\Vert \widetilde{\bC}^{-1} \Vert$, i.e., the relative error is inversely proportional to the smallest singular value of $\widetilde{\bC}$. This may explain a possible failing of approximating matrices, where the smallest singular values tend towards zero. The estimates for the $\H$-Cholesky and the Schur complement for general sparse positive-definite matrices are given in \cite{Bebendorf11}. The approximation errors are proportional to the $\kappa(\widetilde{\bC})$, i.e., matrices with a very large condition number may require a very large $\H$-matrix rank.
\begin{figure}[htbp!]
\centering
\includegraphics[width=0.48\textwidth]{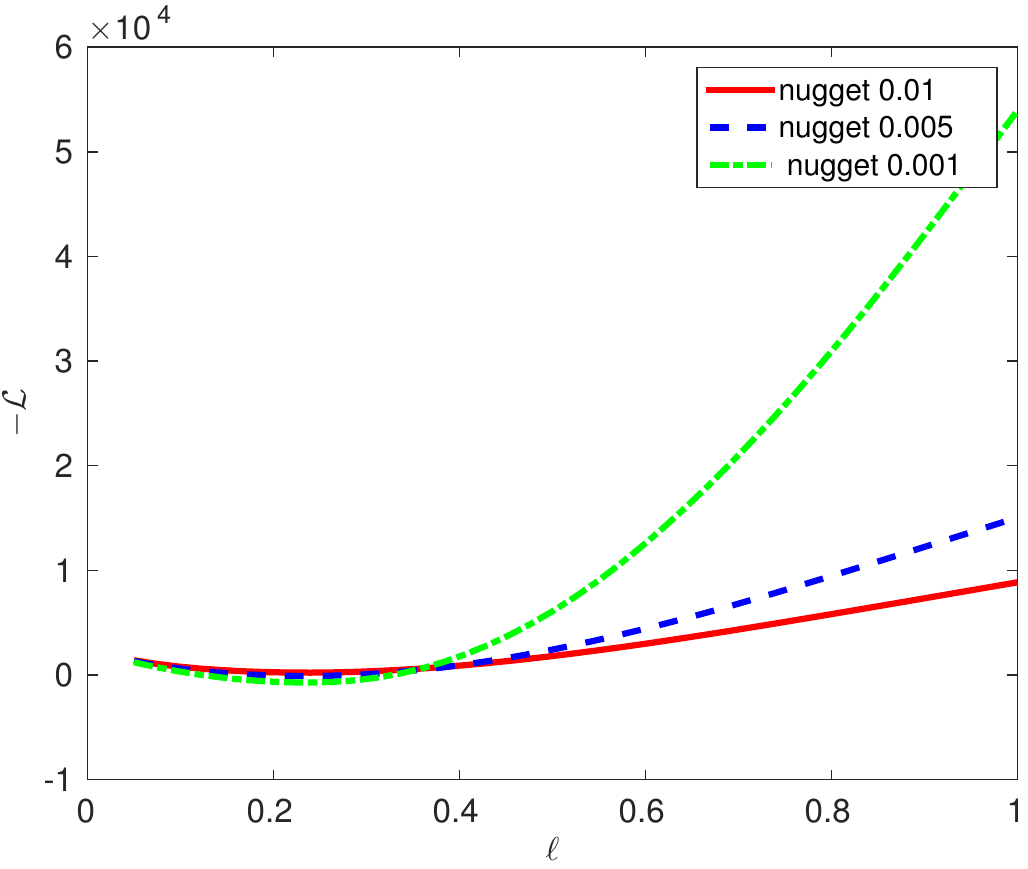}
\includegraphics[width=0.48\textwidth]{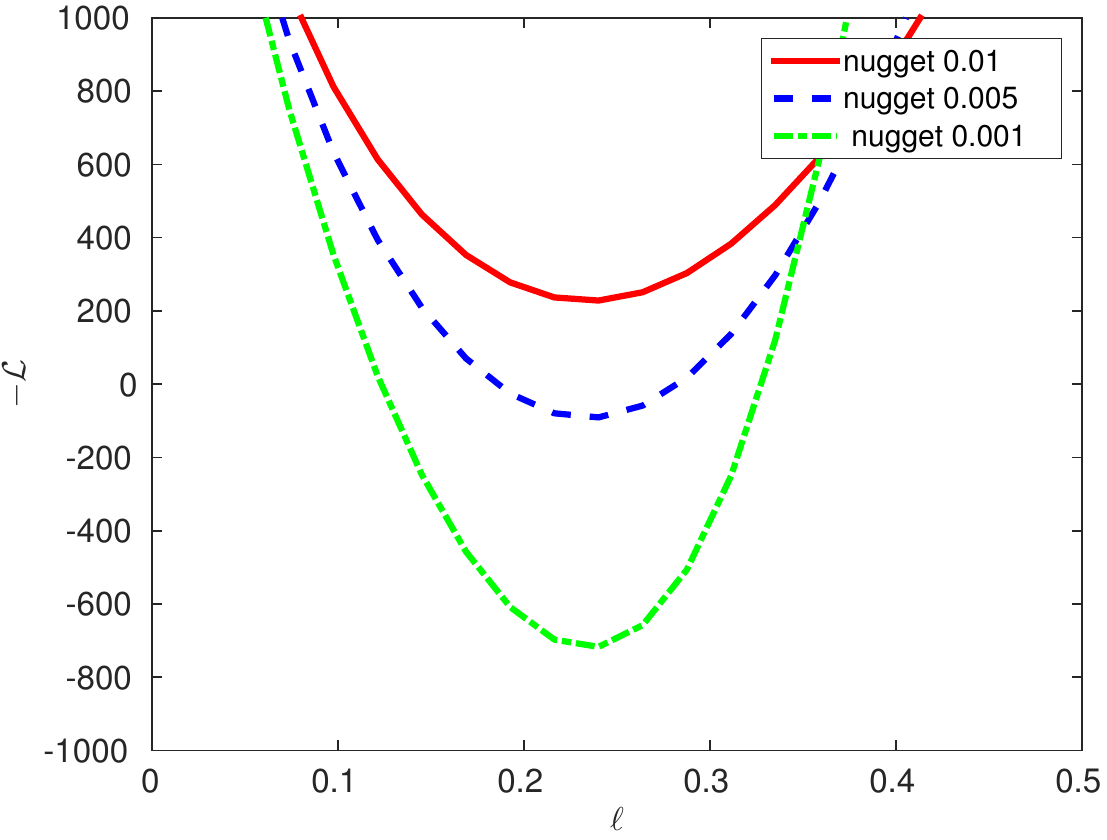}
\caption{
(left) Dependence of the log-likelihood on parameter $\ell$ with nuggets ($\{0.01, 0.005, 0.001 \}$) for Gaussian covariance.
(right) Zoom of the left figure near minimum;
$n=2000$ random locations
, rank $k=14$, $\sigma^2=1$.}
\label{fig:nugget}
\end{figure}
Figure~\ref{fig:nugget} (left) demonstrates three negative log-likelihood functions computed with the nuggets 0.01, 0.005, and 0.001. For this particular example, the behavior of likelihood is preserved, and the minimum does not change (or changes very slightly). Figure~\ref{fig:nugget} (right) is just a zoomed in version of the picture on the left.

\section{Best practices (HLIBCov)}
\label{sec:Best}
In this section, we list our recommendations and warnings.
\begin{enumerate}
\item For practical computations, use adaptive-rank arithmetic since it produces smaller matrices and faster runtime.
\item For the input, it is sufficient to define a file by three columns: both location coordinates ($x$, $y$) and the observed value; no triangles or edges are required.
\item If two locations coincide or are very close to each other, then the matrix will be close to singular or singular. As a result, it will be hard to compute the Cholesky factorization. Our suggested remedy is to improve the quality of the locations by preprocessing the data.
\item By default, the $\H$-Cholesky or $\H-$LU factorizations use a task-based approach employing a DAG (directed
  acyclic graph). For sequential computations this can be turned off to revert to a slightly faster recursive
  implementation by setting
  \begin{quote}
    \texttt{HLIB::CFG::Arith::use\_dag = false}
  \end{quote}
\item By default, HLIBpro uses all available computing cores. To perform computations on 16 cores, use \texttt{HLIB::CFG::set\_nthreads(16)}
at the beginning of the program (after command \texttt{INIT()}).
\item Since HLIBpro is working for 1D, 2D and 3D domains, only very minor changes are required to move from 1D locations to 2D or 3D in HLIBCov. Replace dim$=2$ with dim$=3$ in
  \begin{quote}
    \texttt{TCoordinate coord(vertices, dim);}
  \end{quote}
  then add ``\texttt{ >> z}'' to
  \begin{quote}
    \texttt{in >> x >> y >> z >> v;}
  \end{quote}
\end{enumerate}
The $\H$-matrix data format is a rather complicated data structure (class) in HLIBpro. Therefore, the $\H$-matrix objects (or the pointers on them) are neither the input nor the output parameters. Instead, the input parameters for the HLIBpro C++ routines are: a vector (array) of locations and a vector of observations $\bZ$. The triangulation (a list of triangles/edges) is not needed. The output parameters are either scalar values or a vector; for example, the determinant, the trace, a norm, the result of the matrix-vector product, and an approximation error. 
%
\section{Conclusion}
\label{sec:Conclusion}
We extended functionality of the parallel $\H$-matrix library HLIBpro to infer unknown parameters for applications in spatial statistics.
This new extension allows us to work with large covariance matrices. We approximated the joint multivariate Gaussian likelihood function and found its maxima in the $\H$-matrix format. These maxima were used to estimate the unknown parameters ( $\ell$, $\nu$, and $\sigma^2$) of a covariance model. The new code is parallel, highly efficient, and written in C++ language.
With the $\H$-matrix technique, we reduced the storage cost and the computing cost (Tables~ \ref{table:approx_compare_rank}, \ref{table:approx_compare15}) of the log-likelihood function dramatically, from cubic to almost linear. We demonstrated these advantages in a synthetic example, where we were able to identify the true parameters of the covariance model. We were also able to compute the log-likelihood function for $2{,}000{,}000$ locations in just a few minutes on a desktop machine (Table~\ref{table:approx_compare15}).
The $\H$-matrix technique allowed us to increase the spatial resolution, handle more measurements, consider larger regions, and identify more parameters simultaneously. 

%

\section*{Acknowledgments}
The research reported in this publication was supported by funding from the Alexander von Humboldt foundation (chair of Mathematics for Uncertainty Quantification at RWTH Aachen) and Extreme Computing Research Center (ECRC) at King Abdullah University of Science and Technology (KAUST).
\bibliographystyle{abbrv}

%
%
\begin{appendices}
\section{Admissibility condition}
\label{App:Adm}
Here we give an example of the admissibility criteria \cite{Part1, GH03, Winter}. 
Let
\begin{equation}
\label{eq:green_funcs}
\cov(x, y):=\log \vert x-y\vert, \quad x,y \in \RR^d,
\end{equation}
with singularity at $x=y$. We will introduce a condition, which divides all sub-blocks into admissible and inadmissible. Admissible blocks will be approximated by low-rank matrices.
\begin{defi}
Let $I$ be an index set of all degrees of freedom, i.e. $I=\{1,2,\ldots,n\}$. Denote for each index $i \in I$ corresponding to a basis function $b_i$ the support $\mathcal{G}_i := \supp b_i \subset \RR^d$. 
\end{defi}
Let $\tau,\;\delta \in T_I$ be two clusters (elements of the cluster tree $T_{I}$). Clusters $\tau,\;\delta$ are subsets of $I$, i.e. $\tau,\;\delta \subseteq I$.
We generalise $\mathcal{G}_i$ to clusters $\tau \in T_I$ by setting $\mathcal{G}_{\tau} := \bigcup_{i\in \tau} \mathcal{G}_i$, 
i.e., $\mathcal{G}_{\tau}$ is the minimal subset of $\RR^d$ that contains the supports of all basis functions $b_i$ with $i \in \tau$.

Suppose that $\mathcal{G}_{\tau} \subset \RR^d$ and $\mathcal{G}_{\delta} \subset \RR^d$ are compact and $\chi(x, y)$ is defined for $(x, y) \in \mathcal{G}_{\tau} \times \mathcal{G}_{\delta}$ with $x \neq y$.
The standard assumption on the kernel function in the $\H$-matrix theory is asymptotic smoothness of $\chi(x,y) \in C^{\infty}(\mathcal{G}_{\tau} \times  \mathcal{G}_{\delta})$, i.e, that
\begin{equation*}
 \vert \partial_x^{\alpha}  \partial_y^{\beta} \chi(x,y)\vert \leq C_1 \vert \alpha + \beta \vert ! C_0^{\vert \alpha+\beta \vert} \Vert x-y \Vert^{-\vert \alpha+\beta\vert-\gamma}, \quad \alpha,\,\beta \in  \mathbb{N},
\end{equation*}
holds for constants $C_1$, $C_0$ and $\gamma \in \mathbb{R}$.
This estimation is used to control the error $\epsilon_q$ from the Taylor expansion
\begin{equation*}
\chi(x,y) = \sum_{\alpha\in \mathbb{N}_0^d,\vert \alpha\vert\leq q}(x-x_0)^{\alpha}\frac{1}{\alpha!}\partial^{\alpha}_x\chi(x_0,y) + \epsilon_q.
\end{equation*}

Suppose that $\chi_k(x, y)$ is an approximation of $\chi$ in $\mathcal{G}_{\tau} \times \mathcal{G}_{\delta}$ of the separate form (e.g., Taylor or Lagrange polynomials):
\begin{equation}
\label{eq:separation}
\chi_k (x, y) = \sum_{\nu=1}^k \varphi_{\nu}(x)\psi_{\nu}(y),
\end{equation}
where $k$ is the rank of separation. We are aiming at an approximation of the form (\ref{eq:separation}) such that exponential convergence \begin{equation}
\label{eq:adm_ineq}
\Vert \chi - \chi_k\Vert_{\infty, \mathcal{G}_{\tau}\times \mathcal{G}_{\delta}} \leq \mathcal{O}(\eta ^k)
\end{equation}
holds. 

Let $B_{\tau},\, B_{\delta} \subset \RR^d$ be axis-parallel bounding boxes of the clusters $\tau$ and $\delta$ such that $\mathcal{G}_{\tau} \subset B_{\tau}$ and $\mathcal{G}_{\delta} \subset B_{\delta}$.
\begin{defi}
The \textit{standard admissibility} condition (Adm$_{\eta}$), shown in Fig.~\ref{fig:Hexample_adm} on the left, for two  clusters $\tau$ and $\delta$ is
\begin{equation}
\label{eq:stand_cond}
\min\{\diam(B_{\tau}), \diam(B_{\delta})\} \leq \eta \dist(B_{\tau} , B_{\delta}).
\end{equation}
\end{defi}
Another example is
\begin{equation*}
\label{eq:stand_cond_max}
\max\{\diam(B_{\tau}), \diam(B_{\delta})\} \leq \eta \dist(B_{\tau} , B_{\delta}),
\end{equation*}
where $\eta$ is some positive number.

\begin{defi}
We will say that a pair $(\tau, \delta)$ of clusters $\tau$ and $\delta \in T_I$ is admissible if the condition (\ref{eq:stand_cond}) is satisfied.  The blocks for which condition (\ref{eq:stand_cond}) is true are called admissible blocks.
\end{defi} 
The admissibility condition indicates blocks that allow rank-$k$ approximation and those that do not. Admissible blocks are either very small (and computed exactly) or are approximated by rank-$k$ matrices. All other (inadmissible) blocks are computed as usual.

In order to get a simpler partitioning (see an example in Fig.~\ref{fig:Hexample_adm}, right), we define the \textit{weak admissibility condition} $\mbox{\textit{Adm}}_{W}$ for a pair $(\tau,\delta)$:
\begin{equation}
\label{eq:wead_adm}
   \text{Block}\;\; b = \tau \times \delta \in T_{I \times I} \quad \text{is weak admissible}\Leftrightarrow  (\text{(b is a leaf) or  } \delta \neq \tau ), 
\end{equation}
where $\tau$, $\delta$ are assumed to belong to the same level of $T_{I\times I}$.
 
See more details about derivation of admissibility condition for covariance matrices in \cite{khoromskij2009application}.
\section{Maximum of the log-likelihood function}
\label{App:Appendix}
The C++ code for computing the maximum of the log-likelihood function (\textit{loglikelihood.cc}):
\begin{lstlisting}
double call_compute_max_likelihood(TScalarVector Z, double nu, double covlength, double sigma2,  TBlockClusterTree*  bct, TClusterTree*   ct,  std::vector <double*> vertices, double output[3])
{ gsl_function F;
  int status; iter = 0, max_iter = 200; smy_f_params params ;
  FILE* f1; double size; 
  const gsl_multimin_fminimizer_type *T = gsl_multimin_fminimizer_nmsimplex2;
  gsl_multimin_fminimizer *s = NULL; gsl_vector *ss, *x;
  gsl_multimin_function minex_func;
  params.bct = bct; params.ct = ct; params.Z = Z; params.nu = nu;
  params.covlength=covlength; params.sigma2=sigma2; params.vertices=vertices;
  /* Starting point */
  x = gsl_vector_alloc(3); gsl_vector_set (x, 0, nu);  
  gsl_vector_set (x, 1, covlength);  gsl_vector_set (x, 2, sigma2);
  /* Set initial step sizes to 0.1 */
  ss = gsl_vector_alloc (3);
  gsl_vector_set (ss, 0, 0.02); //for nu 
  gsl_vector_set (ss, 1, 0.04); //for theta
  gsl_vector_set (ss, 2, 0.01);  //for sigma2
  /* Initialize method and iterate */
  minex_func.n = 3; //dimension
  minex_func.f =  &eval_logli;
  minex_func.params = &params;
  s = gsl_multimin_fminimizer_alloc (T, 3); /* optimize in 3-dim space */
  gsl_multimin_fminimizer_set (s, &minex_func, x, ss);
  do{ iter++;
       status = gsl_multimin_fminimizer_iterate(s);
       if (status) break;
       size = gsl_multimin_fminimizer_size (s);  //for stopping criteria
       status = gsl_multimin_test_size (size, 1e-5);  
       if (status == GSL_SUCCESS) printf ("converged to minimum at \n");}}
  while (status == GSL_CONTINUE && iter < max_iter);
  output[0]= gsl_vector_get(s->x, 0); //nu
  output[1]= gsl_vector_get(s->x, 1); //theta
  output[2]= gsl_vector_get(s->x, 2); //sigma2
  gsl_vector_free(x);   gsl_vector_free(ss); gsl_multimin_fminimizer_free (s);
  return status; }
 \end{lstlisting}
%
%
Below we list  the C++ code, which computes the value of the log-likelihood for given parameters (\textit{loglikelihood.cc}): 
\begin{lstlisting}
double eval_logli (const gsl_vector *sol, void* p)
{   pmy_f_params params  ;
    double nu = gsl_vector_get(sol, 0);
    double length = gsl_vector_get(sol, 1);
    double sigma2 = gsl_vector_get(sol, 2);
    unique_ptr< TProgressBar >  progress( verbose(2) ? new TConsoleProgressBar : nullptr );
    params = (pmy_f_params)p;
    TScalarVector rhs= (params->Z);
    TBlockClusterTree* bct = (params->bct);  TClusterTree* ct = (params->ct);
    vector< double * > vertices= (params->vertices);
    double err2=0.0, nugget = 1.0e-4,  s = 0.0;
    auto                acc = fixed_prec( 1e-5 ); int  dim = 2, N = 0;
    TCovCoeffFn  coefffn(length,nu,sigma2,nugget,vertices,ct->perm_i2e(),ct->perm_i2e());
    TACAPlus< real_t >          aca( & coefffn );
    TDenseMatBuilder< real_t >  h_builder( & coefffn, & aca );
    // enable coarsening during construction
    h_builder.set_coarsening( false );
    auto  A = h_builder.build( bct, acc, progress.get() );
    N=A->cols();
    auto  A_copy  = A->copy();
    auto  options = fac_options_t( progress.get() );
    options.eval = point_wise;   //! Extreme important
    auto  A_inv = ldl_inv( A_copy.get(), acc, options );
    for ( int  i = 0; i < N; ++i ) {
            const auto  v = A_copy->entry( i, i );
            s = s + log(v);}// for
     TStopCriterion  sstop( 150, 1e-6, 0.0 );
     TCG             solver( sstop );
     TSolverInfo  sinfo( false, verbose( 4 ) );
     auto            solu = A->row_vector();
     solver.solve( A.get(), solu.get(), & rhs, A_inv.get(), & sinfo );
     auto  dotp = re( rhs.dot( solu.get() ) );
     auto  LL = 0.5*N*log(2*Math::pi<double>())+0.5*s+0.5*dotp;}
\end{lstlisting}
\paragraph{Rank-$k$ Adaptive Cross Approximation (ACA):}
An $\H$-matrix contains many sub-blocks, which can be well approximated by low-rank matrices. 
These low-rank approximations can be computed accurately by truncated singular value decomposition (SVD), but it is very slow.
HLIBpro uses the Adaptive Cross Approximation method (ACA) \cite{TyrtyshACA} and its improved modifications such as ACA+ and HACA \cite{ACA, ACA2, HACA}.

\begin{rem}
Further optimization of the ACA algorithm can be achieved by a recompression using low-rank SVD. If we suppose that a
factorization of the matrix $\R=\A\B^\top$, $\A\in \RR^{p\times K}$, $\B\in \RR^{q\times K}$, is found by ACA and that
the actual rank of $\R$ is $k$, $k<K$. Then we can apply the low-rank SVD algorithm to compute $\R=\U\Sigma\V^\top$ in
$\mathcal{O}((p+q)K^2+K^3)$ time.
\end{rem}

\end{appendices}

\end{document}